%% file: apssamp.tex
\documentclass[aps,prc,twocolumn,superscriptaddress,nofootinbib]{revtex4-2}

\usepackage{graphicx}% Include figure files
\usepackage{dcolumn}% Align table columns on decimal point
\usepackage{bm}% bold math
\usepackage{color}
\usepackage[makeroom]{cancel}
\usepackage[colorlinks=true, linkcolor=black, urlcolor=blue, citecolor=blue]{hyperref}
\usepackage{cleveref}
\usepackage{subcaption}
\usepackage{subfiles}

\begin{document}

\preprint{APS/123-QED}

\title{The MUSE Beamline Calorimeter}% Force line breaks with \\
%Perhaps use alternate title?
%\title{Calorimeter in the MUon Proton Scattering Experiment (MUSE)}% Force line breaks with \\

\input{authorlist.tex}

\begin{abstract}
The MUon Scattering Experiment (MUSE) was motivated by the proton radius puzzle arising from the discrepancy between muonic hydrogen spectroscopy and  electron-proton measurements. 
The MUSE physics goals also include testing lepton universality, precisely measuring two-photon exchange contribution, and testing radiative corrections. 
% MUSE addresses these physics goals through high-precision cross sections for electron-proton and muon-proton scattering, measured simultaneously in a mixed beam.
MUSE addresses these physics goals through simultaneous measurement of high precision cross sections for electron-proton and muon-proton scattering using a mixed-species beam.
The experiment will run at both positive and negative beam polarities. 
%In addition to measuring the charge radius of the proton, MUSE will also test lepton universality and two photon exchange by comparing the cross section ratios of lepton scattering. 
Measuring precise cross sections requires understanding both the incident beam energy and the radiative corrections. 
For this purpose, a lead-glass calorimeter was installed at the end of the beam line in the MUSE detector system. 
In this article we discuss the detector specifications, calibration and performance. 
We demonstrate that the detector performance is well reproduced by simulation, and meets experimental requirements.
% We will show that the detector performance is well reproduced by simulation, and meets experimental requirements. 

\end{abstract}

%\keywords{Suggested keywords}%Use showkeys class option if keyword display desired
\maketitle

%\tableofcontents

\section{\label{sec:intro}Introduction}

%\textit{\color{red} At times reading through, I thought we were under-referenced; more citations may be needed of earlier work.}

The Proton Radius Puzzle was the observation of a significantly smaller proton charge radius determined in muonic hydrogen spectroscopy than had been determined from atomic hydrogen spectroscopy and electron proton scattering \cite{muonic_hydrogen_2010, muonic_hydrogen_2013}.
%larger than $5\sigma$ difference from the radius
This observation led to strong interest in the proton radius and related physics of potential new forces, lepton universality, and radiative corrections, including two-photon exchange and polarizibility. 
Possible explanations for the puzzle  generally fall into the categories of new forces, novel aspects of conventional physics, or issues in experimental extractions of the radius.
% While there are new electronic measurements of the radius in agreement with the muonic spectroscopy result, the overall agreement among the new results suggests that there are ongoing experimental issues in determining the radius.
%such as the issues in mis-treatment of the radiative corrections in the older scattering measurements, possible theory improvement in spectroscopy, and potential new physics, however, the puzzle remains unsolved. There have been new experiments measuring the proton radius using atomic spectroscopy and electron proton scattering, but the results are contradictory. Re-analysis of the older data is also not able to conclude the puzzle. 
While there are new electronic measurements of the radius in agreement with the muonic spectroscopy result, the lack of overall agreement among the new results suggests that there are ongoing experimental issues in determining the radius \cite{Jan_review}.

The MUSE experiment, motivated by a lack of high-precision muon-scattering data that addresses these physics issues, is working towards a sub-percent level, high-precision experiment that simultaneously measures elastic electron-proton scattering and muon-proton scattering.
%The experiment alternates between positive and negative beam polarities. 
%MUSE will provide more data to resolve the proton radius puzzle. In addition, MUSE will test lepton universality and two photon exchange by comparing the cross section ratios of the lepton scattering.
MUSE is located at the Paul Scherrer Insititute in Villigen, Switzerland.
The High-Intensity Proton Accelerator facility uses a  cyclotron to produce an approximately 2 mA, 590 MeV proton beam. 
The beam passes through the graphite-wheel M target to produce electrons, muons and pions that are transported by the PiM1 channel to MUSE.
MUSE has run at three momentum settings, of 115 MeV/$c$, 160 MeV/$c$ and 210 MeV/$c$, at selectable positive or negative beam polarity \cite{beam_paper}. 

\begin{figure}[h]
\centering
\includegraphics[width=0.45\textwidth, trim=1.2cm 5cm 1cm 4.5cm, clip]{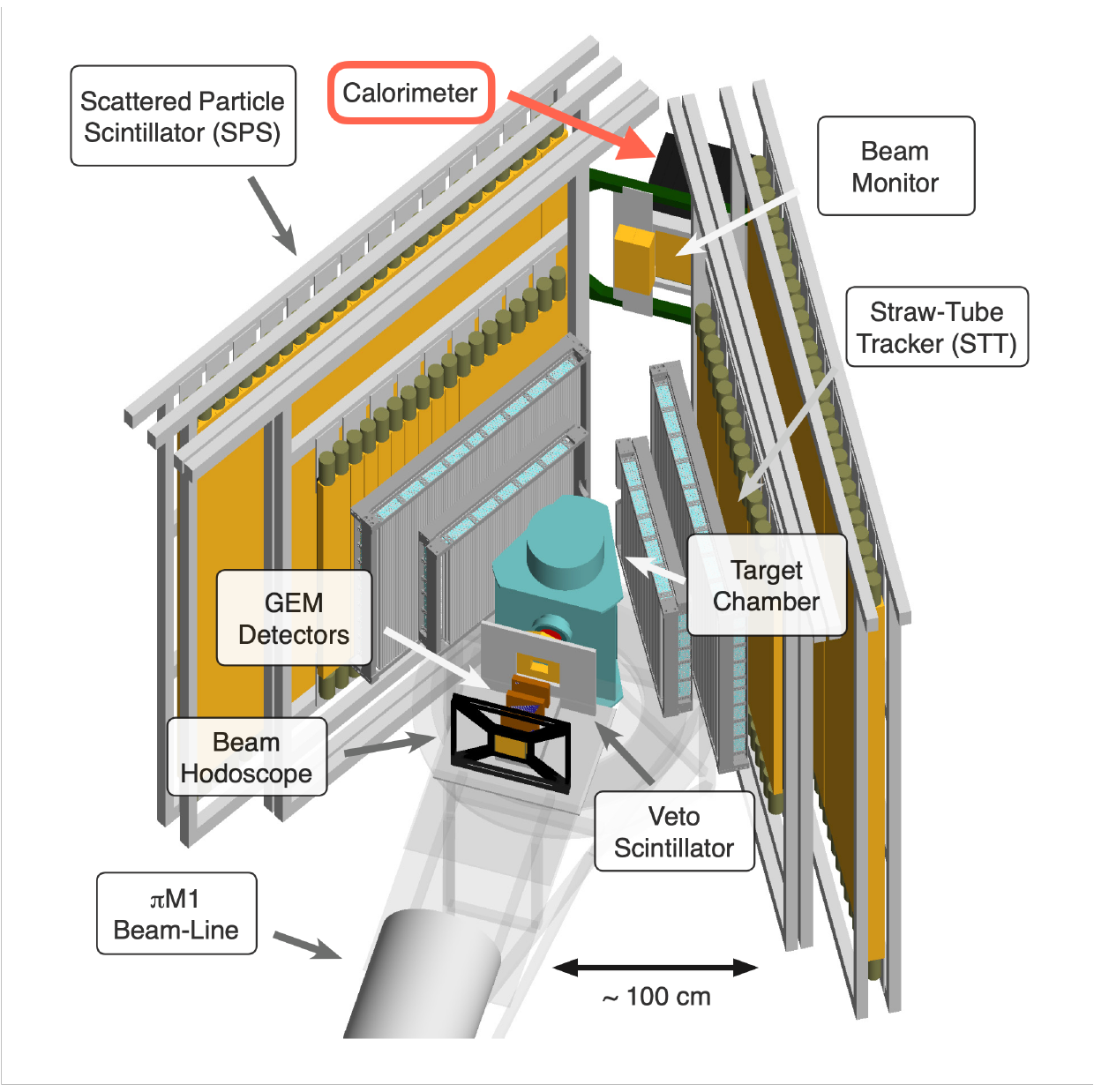}
\caption{\label{fig:muse_setup} Geant4 drawing of the MUSE system, showing the calorimeter at the downstream (upper right) end of the detector system. } 
\end{figure}

MUSE comprises beamline and scattering detectors. 
Figure~\ref{fig:muse_setup} is a drawing of the full detector setup. % generated by the Geant4 simulation. 
The first two detectors in the beam are a two-plane plastic scintillator Beam Hodoscope (BH) \cite{sipm_paper} and a four-plane Gas Electron Multiplier (GEM) detector. 
These two detectors provide incoming particle timing, species identification, and tracking.
Downstream of the GEM detector are the veto scintillator detector (VETO) and the trapezoidal target chamber \cite{Target_Paper}. 
The annular VETO detector rejects events with beam particle trajectories that hit the target chamber walls, such as decays in flight and beam halo. 
On both sides of the target chamber, the Straw-Tube Tracker (STT) and Scattered Particle Scintillator (SPS) are responsible for scattered event tracking and timing. 
%These detectors cover a scattering angle range of about $\theta = 20^{\circ}$ -- $100^{\circ}$, corresponding to a $Q^2$ range of 0.0016 to 0.0820 (0.0799) $\rm{GeV^2}$ for electrons (muons).
Downstream of the target chamber the Beam Monitor (BM) and calorimeter (CALO) help monitor the beam, measure its energy, and identify events with high energy forward-going photons. 

Extracting  precise Born cross sections and the proton radius from the experimental scattering cross sections requires radiative corrections to account for the contribution of higher-order processes to the scattering cross section.
The leading-order correction with additional particles in the final state is the Bremsstrahlung 
%(Brem) 
correction, in which the charged leptons or protons emit real photons when accelerated.
This correction generates a radiative tail, a continuous distribution in momentum for the nominally two-body elastic scattering $ep \to ep$ reaction.
This correction is most significant for electrons, as it is suppressed for muons and pions due to their higher mass.
%To measure scattering cross sections precisely, understanding of both the incident beam energy and the radiative correction are required.

Most electron-scattering experiments use magnetic spectrometers, and the measured spectrum and limited momentum  acceptance of the experiments allow the Bremsstrahlung correction to be calculated with 
%\textcolor{red}{a} 
good precision.
MUSE, however, runs 
%at low beam momenta 
with a large solid angle, non-magnetic spectrometer.
As a result, MUSE measurements integrate over a wide range of outgoing electron momenta, including most of the radiative tail, to determine the cross section.
The experimental uncertainty is sensitive to the limits of integration over the radiative tail, determined by detector thresholds on particle energy loss. 
%To limit the radiative corrections and reduce the sensitivity to them, limiting experimental uncertainties, MUSE employs two strategies.
MUSE employs two strategies that allow radiative corrections to be studied with the experimental data, with the goal of limiting the associated uncertainty from these corrections.
One strategy is to use a low detector hardware threshold along with a software energy loss cut in the analysis.
This procedure allows the cut and its uncertainties to be checked with simulations.
A second strategy is to suppress the Bremsstrahlung correction by removing events from the analysis with high-energy, forward-going photons.
This limits the initial-state radiation, largely removing the low-energy tail of the momentum distribution, due to the dominance of the initial-state over the final-state Bremsstrahlung correction -- the initial-state raidation moves the vertex to lower $Q^2$ and thus higher cross section.

\begin{figure}[h]
\centering
\includegraphics[width=0.45\textwidth, trim=3.2cm 2.5cm 3cm 3cm, clip]{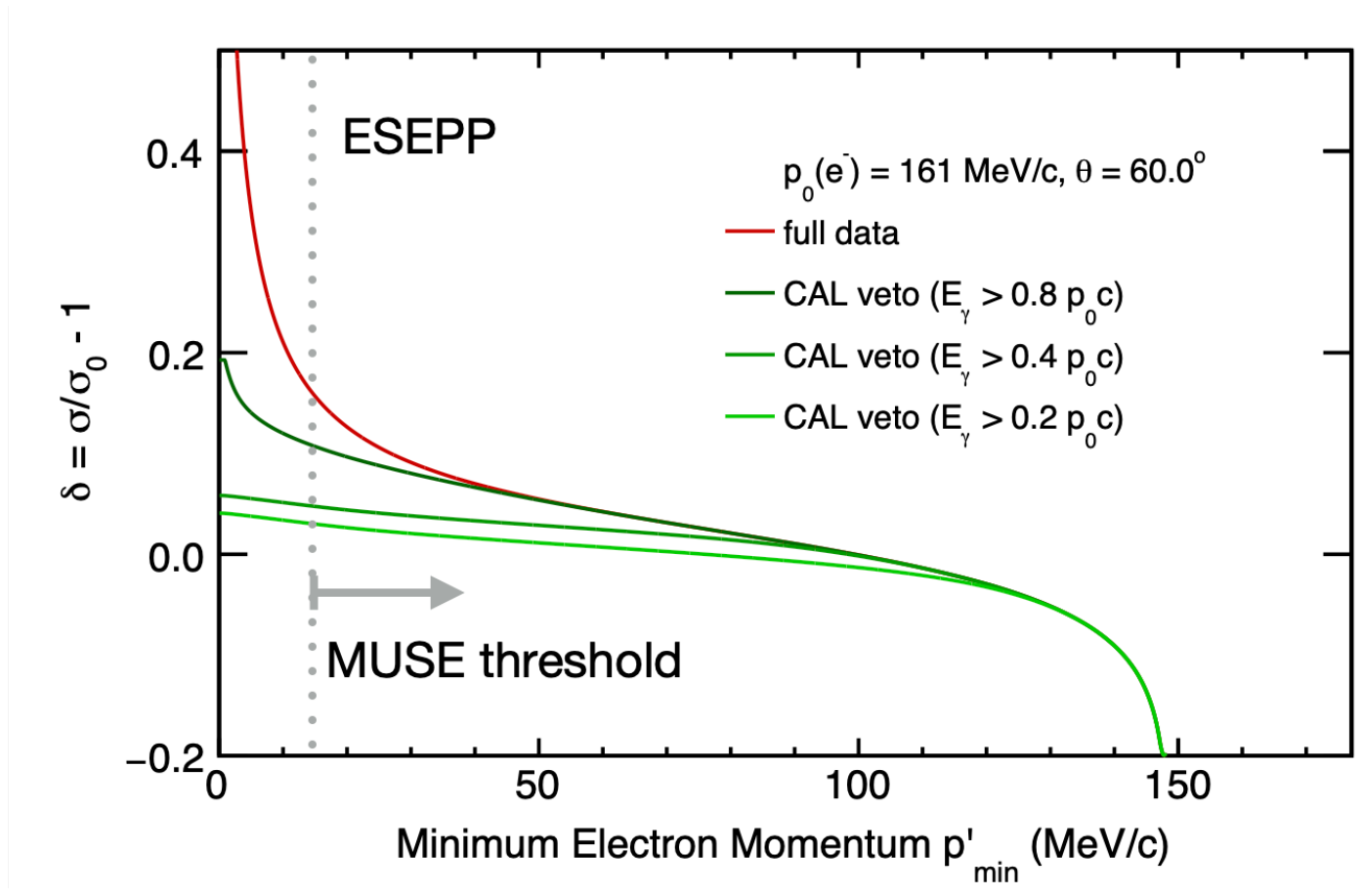}
\caption{\label{fig:correction_and_cuts} Radiative corrections for $ep$ scattering as a function of the minimum electron momentum at one MUSE kinematic setting, calculated using the ESEPP event generator \cite{esepp_paper}. 
The radiative correction factor $\delta$ indicates the  difference between the experimental and Born cross sections:  $\sigma/\sigma_0 = 1 + \delta$.
See Ref.~\cite{Li:2023sxf} for details.} 
\end{figure}

Removing events with high-energy photons is done with the lead-glass calorimeter installed at the most downstream position of the experiment. Figure~\ref{fig:correction_and_cuts}, reproduced from Ref.~\cite{Li:2023sxf}, shows an example of the dependence of the radiative correction on the minimum electron momentum in the integration. 
The red curve represents the radiative correction for $ep$ scattering when there is no suppression of high-energy photons. 
The green curves are the radiative correction  including calorimeter energy cuts in the data analysis.
With the photon energy cuts, the corrections are less 
%steep 
sensitive to the electron momentum threshold in the MUSE threshold region, where integration begins.
Thus, the sensitivity and the resulting uncertainties are  reduced. 
In particular, with a photon energy cut of greater than 40\% of the beam momentum, the correction is small and nearly linear in the region of the cut, limiting this important experimental systematic correction and uncertainty  \cite{Li:2023sxf}.
The variation of cross section with these radiative corrections cuts is a strong test for MUSE that can demonstrate the quality of the applied radiative corrections, and that they are well understood.

%{\it\color{red} Draft calo spec paragraph (needs numbers):}
The MUSE calorimeter specifications were developed based on Geant4 simulations with ESEPP radiative corrections \cite{esepp_paper} shown in Fig.~\ref{fig:correction_and_cuts}.
%With the calorimeter used to veto events, the cross section is then increased by detector inefficiency.
The cross section can be increased or decreased by offsets in the calorimeter light output calibration and by detector resolution, depending on the slope of the flux and the variation in resolution near threshold.
With the goal of limiting changes in the absolute cross section from these resolution and offsets to the 0.1\% level, we observed in the simulation that the calorimeter light output scale needed to be determined to approximately 2 MeV in the cut region, and corrections could be performed at an acceptable level for a resolution near threshold of approximately 5\%/$\sqrt{E}$, with $E$ in GeV.

In the following sections, we discuss the hardware details, calibration procedures and performance of the calorimeter. 
We show that detector simulations agree well with the experimental data.
 %showing our understanding of the detector performance will also be discussed.

\section{\label{sec:hardware}Detector Description}

\begin{figure}[h]
\centering
\includegraphics[width=0.45\textwidth, trim={2.1cm 3.5cm 1.4cm 6cm}, clip]{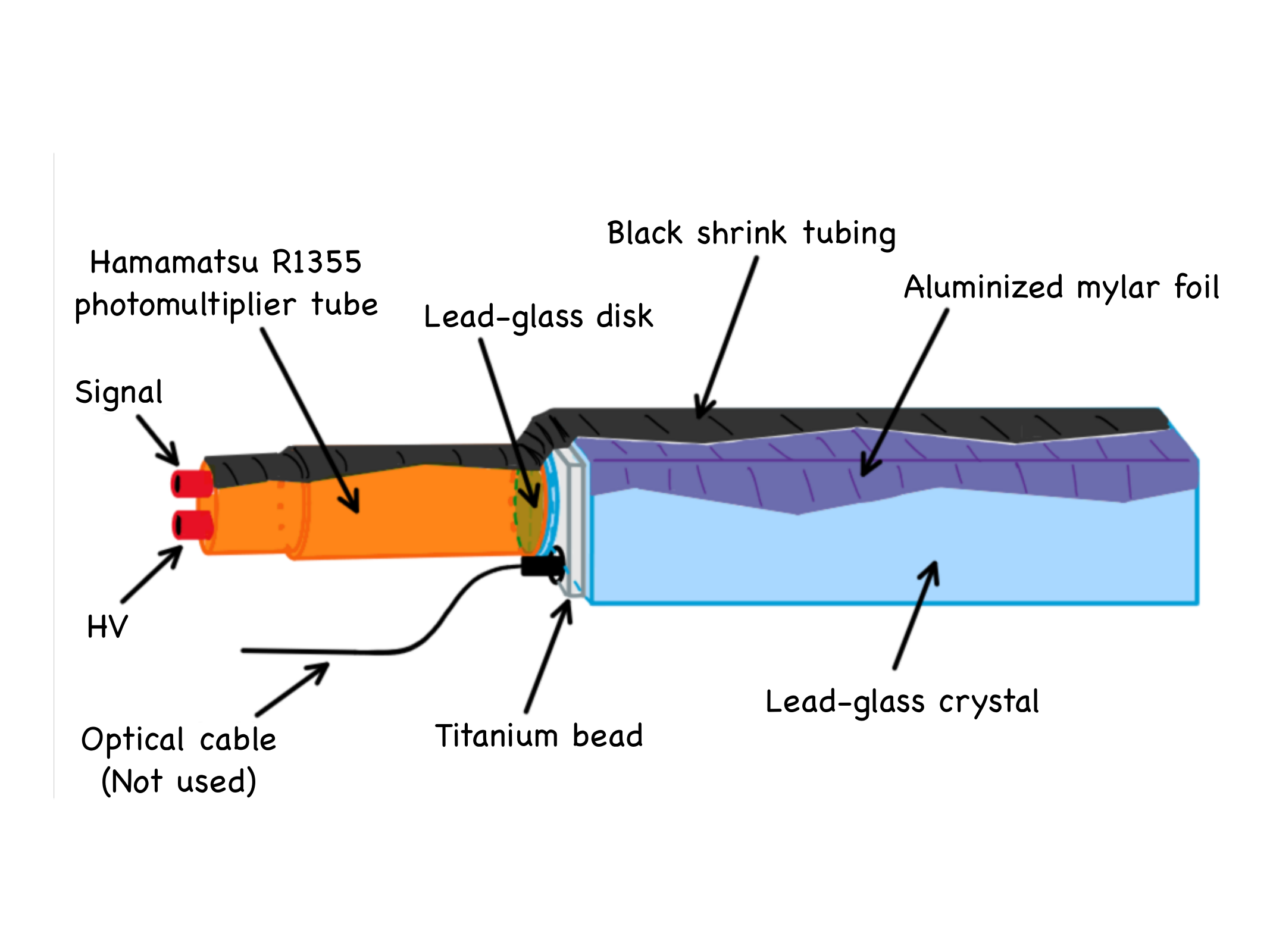}
\caption{\label{fig:calo_schematic} Schematic drawing of a calorimeter crystal. 
Each crystal is 4 cm $\times$ 4 cm $\times$ 30 cm 
%(24 radiation lengths) 
in dimension. 
%The longest side is placed along the beam direction.}
The PMT shown to the left is at the downstream end of the crystal \cite{calo_original, Christopher_report}.}
\end{figure}

The calorimeter detector consists of an 8$\times$8 array of SF5 lead-glass crystals\footnote{Each crystal has optical fibers that can input calibration test signals. These unused optical cables were sealed into a light-tight box located on top of the crystal array.}, on loan from the A2 experiment @ MAMI~\cite{calo_original}.
The radiation length and Molière radius of lead-glass are 1.265~cm and 2.578~cm, respectively~\cite{pdg_2023}. 
Figure~\ref{fig:calo_schematic} shows a schematic drawing of a calorimeter crystal and the associated photomultiplier.
The crystals are 24 radiation lengths long along the beam direction.
Each lead-glass crystal is individually wrapped with aluminized Mylar foil and black shrink tubing for optical isolation. 
The detector signals are read out by Hamamatsu R1355 photo-multiplier tubes (PMTs), which produce a negative anode signal.

When particles directly hit the center of a crystal, that crystal absorbs most of the particle's kinetic energy, with a small fraction of the energy spreading to neighboring crystals.
However, particles can hit edges and corners of the crystals, depositing significant energy in the neighboring crystals. 
%\textcolor{blue}{Hence, when determining the energy of an event with maximum energy, a sum of energies deposited in 8 surrounding neighbor crystals is taken into account.}
Hence, 
%when summing energies for each event, 
most of the energy of an incoming particle is easily captured by summing the energy of the crystal with maximum energy with the energies of the 8 surrounding neighbors.
The energy response and analysis will be discussed more in Sec.~\ref{sec:performance}. 

\begin{figure}[h]
\centering
\includegraphics[width=0.45\textwidth,  trim={0cm 6.2cm 0cm 6.2cm}, clip]{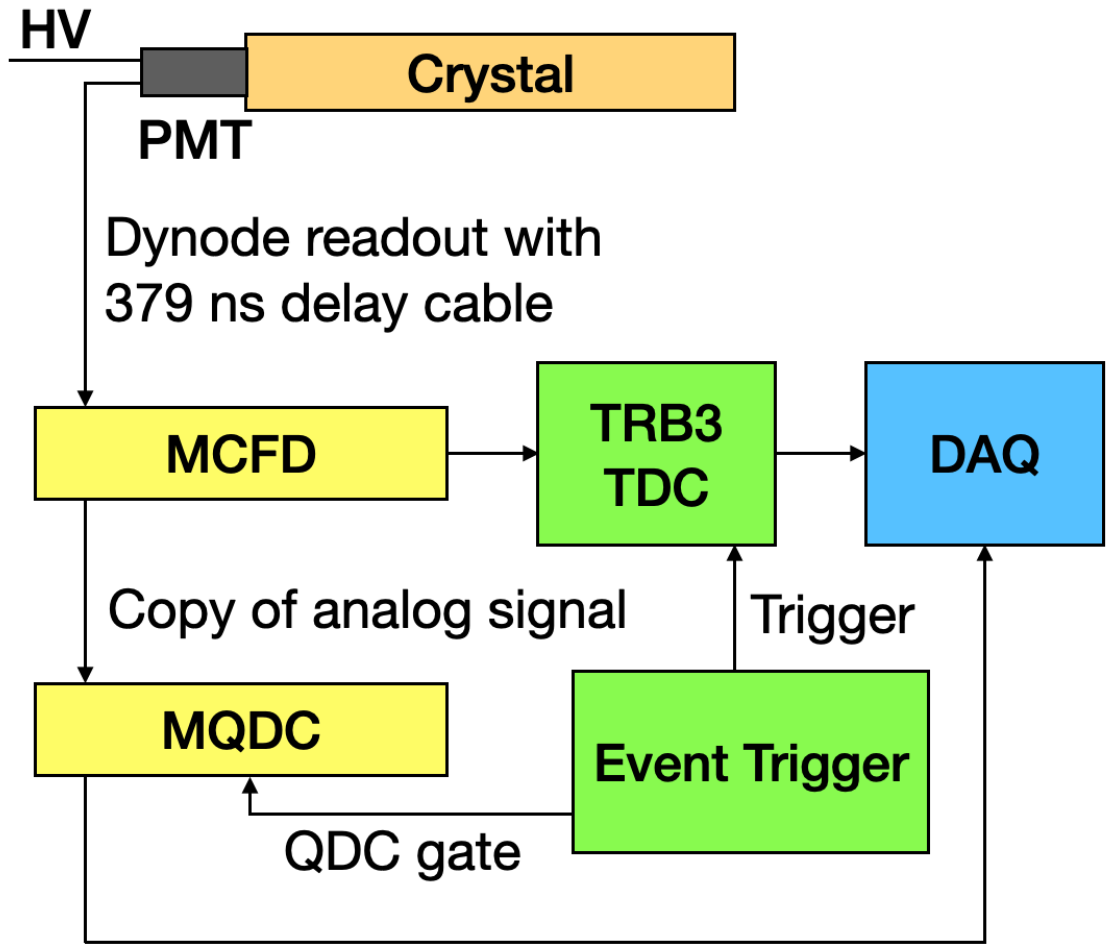}
\caption{\label{fig:calo_readout} Schematic drawing of the calorimeter readout.} 
\end{figure}

\begin{figure}[h]
\centering
\includegraphics[width=0.45\textwidth, trim={0cm 3.8cm 0cm 4cm}, clip]{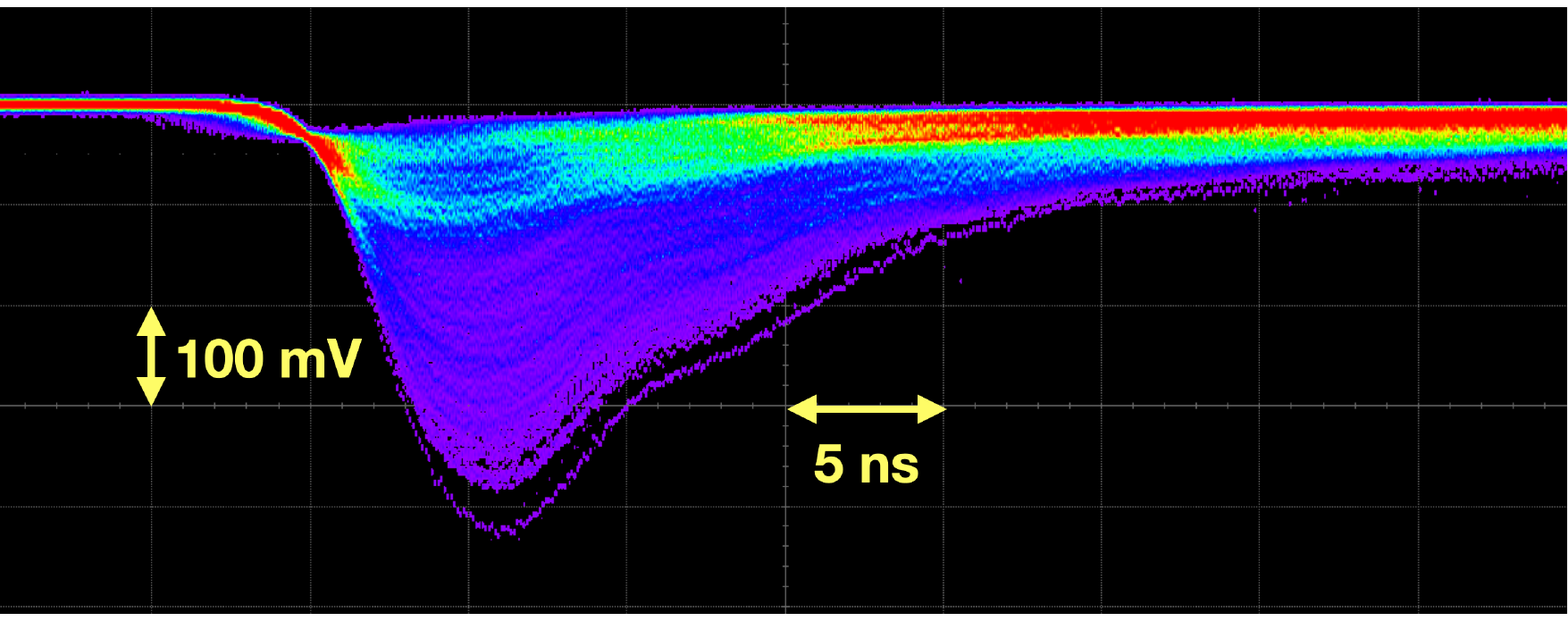}
\caption{\label{fig:calo_scope} Calorimeter signals from a mixed, negatively charged particle beam at 210 MeV/$c$, after the 379 ns long delay.} 
\end{figure}

Figure~\ref{fig:calo_readout} shows a schematic drawing of the readout electronics.
The signals propagate through delay lines before entering the Mesytec MCFD-16 constant fraction discriminators (CFDs) \cite{mcfd}.
Figure~\ref{fig:calo_scope} shows typical crystal signals in an oscilloscope, after the delay, before being input to the CFDs.
Data are for a 210 MeV/$c$ negative polarity beam, comprised of approximately 30\% $e$'s, 5\% $\mu$'s, and 65\% $\pi$'s, with $e$'s ($\pi$'s) generating the largest (smallest) signals.
The CFDs generate logical timing signals that are digitized by TRB3 TDCs \cite{trb3}.
The CFDs also send copies of the analog signals to  Mesytec MQDC-32 Charge to Digital Converters (QDCs) \cite{mqdc} for integration with 12-bit precision.
While the QDC information is sufficient to determine light output in the detector for each event, the few MHz beam rate leads to light output from randomly coincident beam particles. 
The TDC information helps to distinguish in-time clusters from a scattering event from clusters from randomly coincident beam particles that are also read out in the same event.

\begin{figure}[h]
\centering
\includegraphics[width=0.45\textwidth, trim={4cm 0cm 4cm 0cm}, clip]{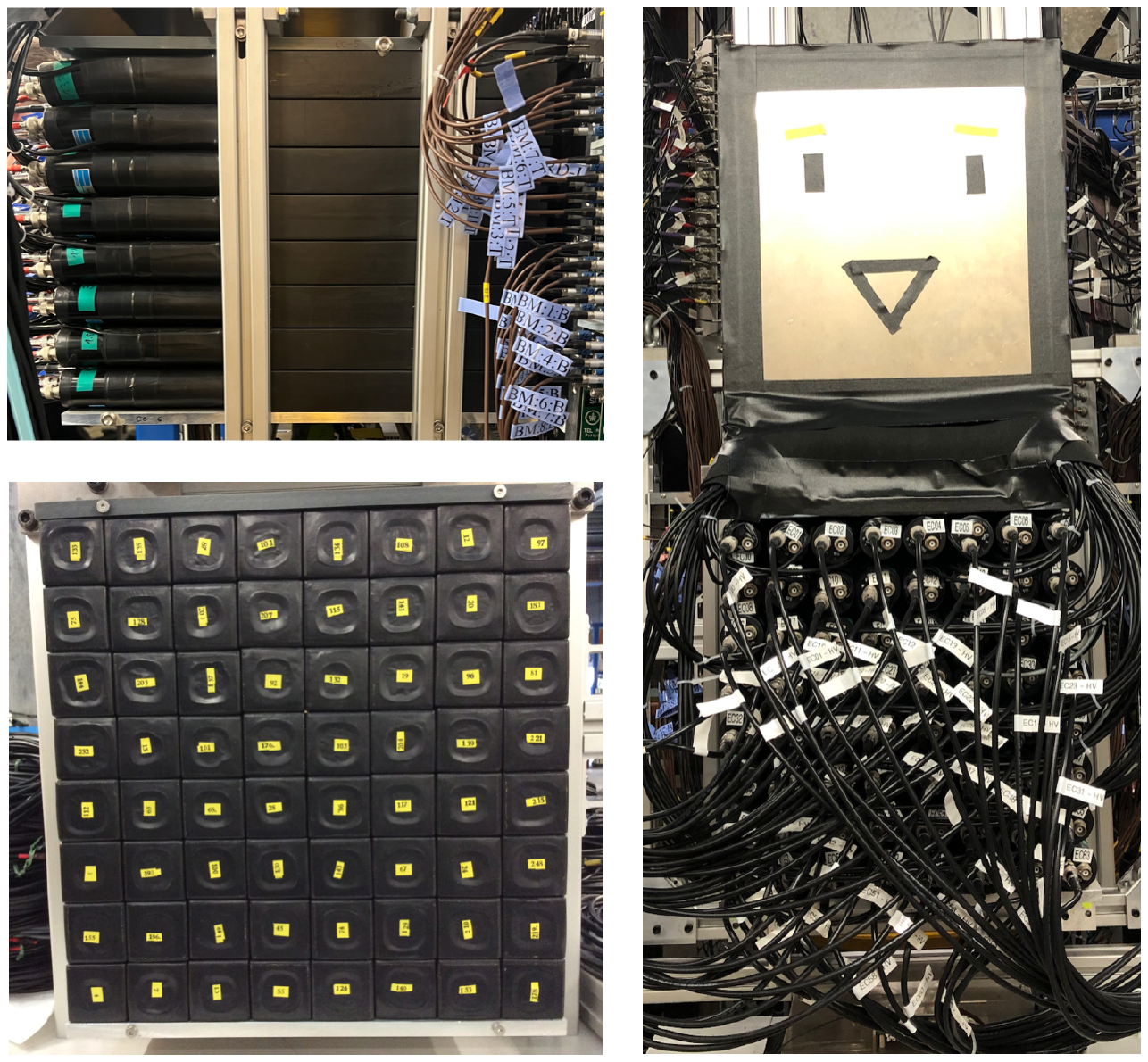}
\caption{\label{fig:calo_pic} Photographs of the calorimeter from three  perspectives. \textbf{Top Left:} Side view from beam left. \textbf{Bottom Left:} Looking downstream towards the upstream face of the calorimeter. \textbf{Right:}
Looking upstream at the full detector including the light-tight box on top of the crystals and the PMTs.} 
\end{figure}

\begin{figure}[h]
\centering
\includegraphics[width=0.4\textwidth]{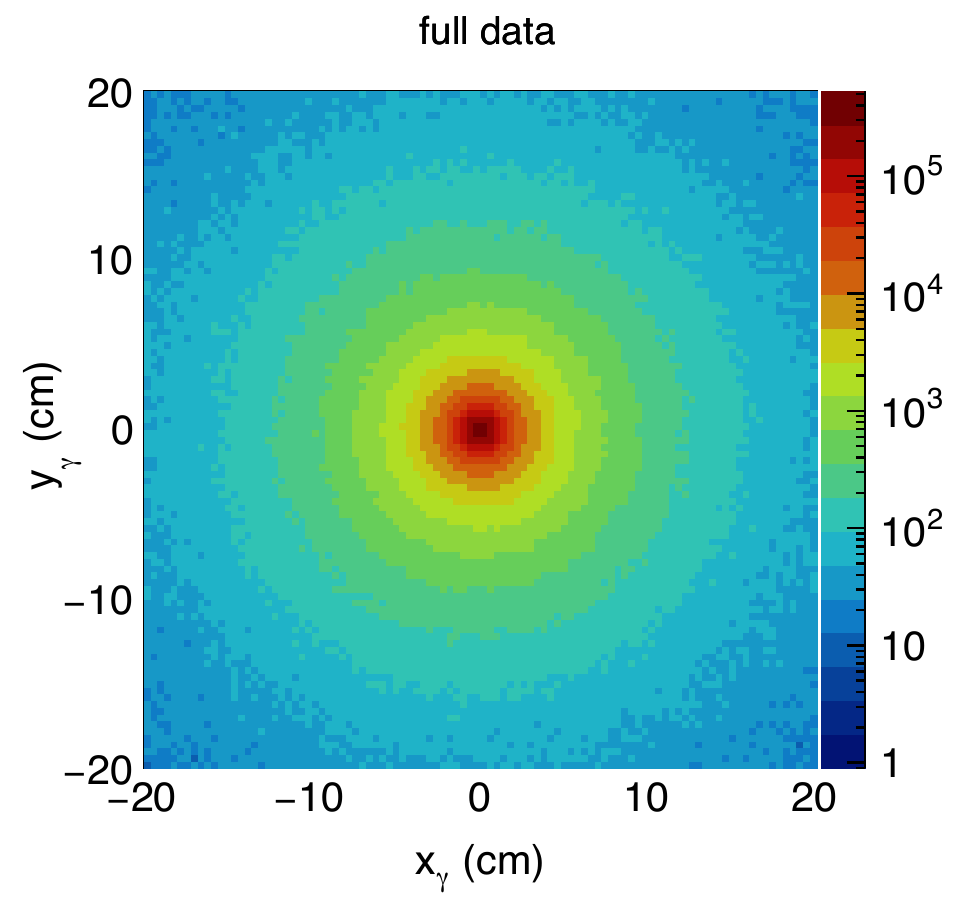}
\includegraphics[width=0.4\textwidth]{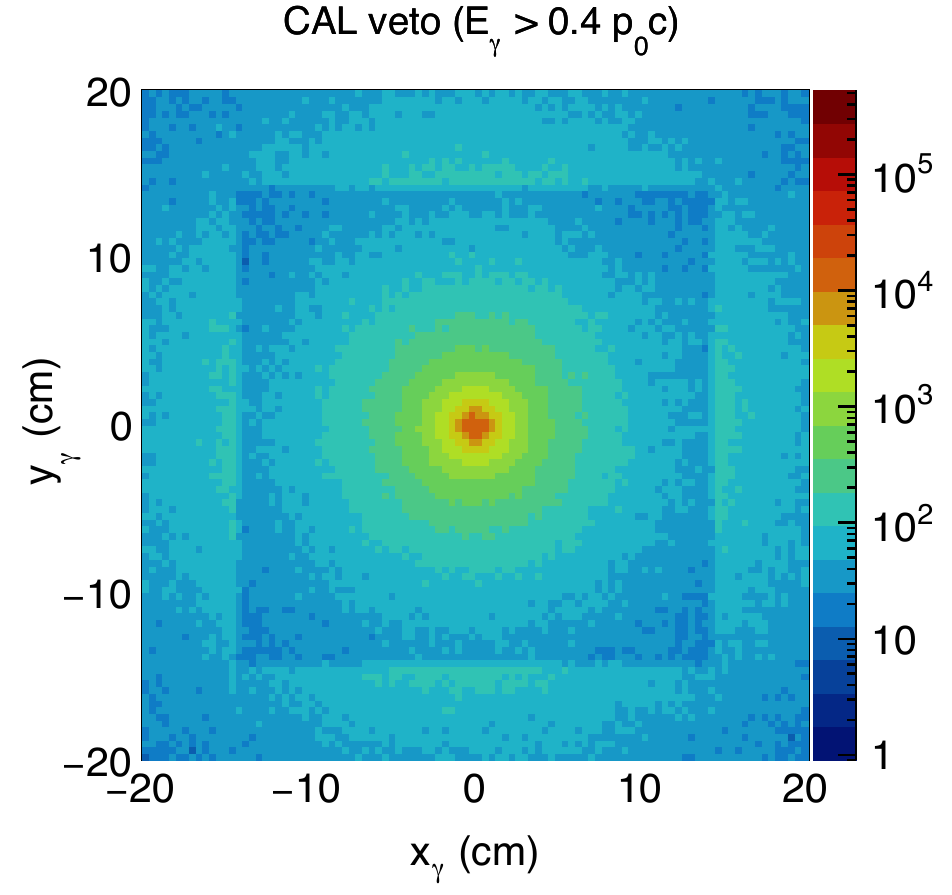}
\caption{\label{fig:calo_photon} Simulated photon distribution at the front face of the calorimeter.
\textbf{Top:} All photons.
\textbf{Bottom:} Distribution of photons after applying a calorimeter energy cut. The central area with less counts compared to the top plot reflects the size of the detector.} 
\end{figure}

Figure~\ref{fig:calo_pic} shows  pictures of the calorimeter detector viewed from different directions. 
The front face of the calorimeter is located 138.5 cm from the center of the target. The detector covers an area of about 33 cm by 33 cm, or an angular range of about $\pm$6.8$^{\circ}$ in horizontal and vertical directions, which is sufficient to capture most of the forward-going photons.
%\textit{\color{red} I think the following sentences maybe belong in analysis / simulation section?}
Figure~\ref{fig:calo_photon} shows the simulated photon distribution at the front face of the calorimeter before and after the calorimeter cuts. 
With this calorimeter design, most of the high-energy-photon events will be removed in the analysis, while the the lower energy photon events will be retained.

%The MUSE calorimeter reads out both TDC and QDC information. While the QDC information gives energy deposited in the detector by each event, the TDC helps to distinguish multiple clusters from the same event. The signal of the detector is a negative anode signal, which is intentionally delayed and sent to the Mesytec CFD to read out timing information. A copy of the analog signal is sent from the MCFD to the Mesytec QDC for signal integration. The MQDC receives QDC gate from the event trigger and integrates the signals. A diagram showing the readout logic is shown in Figure~\ref{fig:calo_readout}.

\section{\label{sec:calibration}Calibration Procedures}

The goal of the calibration procedure is to convert the signal sizes in each detector channel, which are proportional to the light generated by electrons in the crystals, to an energy, so that the incident electron energies can be determined.
The calibration requires both gain matching the calorimeter crystals and determining the light output scale. 
Gain is the proportionality between the output QDC values and the particle signal size in the detector. 
We calibrate the gain for each channel to ensure the detector light output derived is consistent for all channels. 
It is convenient to use both cosmic rays and beam particles for this purpose.
Cosmic rays are always available and illuminate the calorimeter crystals more uniformly than does the beam.
However, light from cosmic rays is not collected with the same efficiency as for beam particles due to the different trajectories of the particles, so determining an absolute light output scale is more straightforward with beam particles of known energy.
We focus on the response of the calorimeter to electrons in the beam, which generate similar signals in the calorimeter to photons of the same energy.
Thus, we initially gain match with cosmics, and then we calibrate the absolute energy scale with beam.
%Instead we use energy sums with beam particles to calibrate the QDC to energy conversion of the gain matched crystals.

\begin{figure}[h]
\centering
\includegraphics[width=0.45\textwidth, trim={4.5cm 0.8cm 4.5cm 1.5cm}, clip]{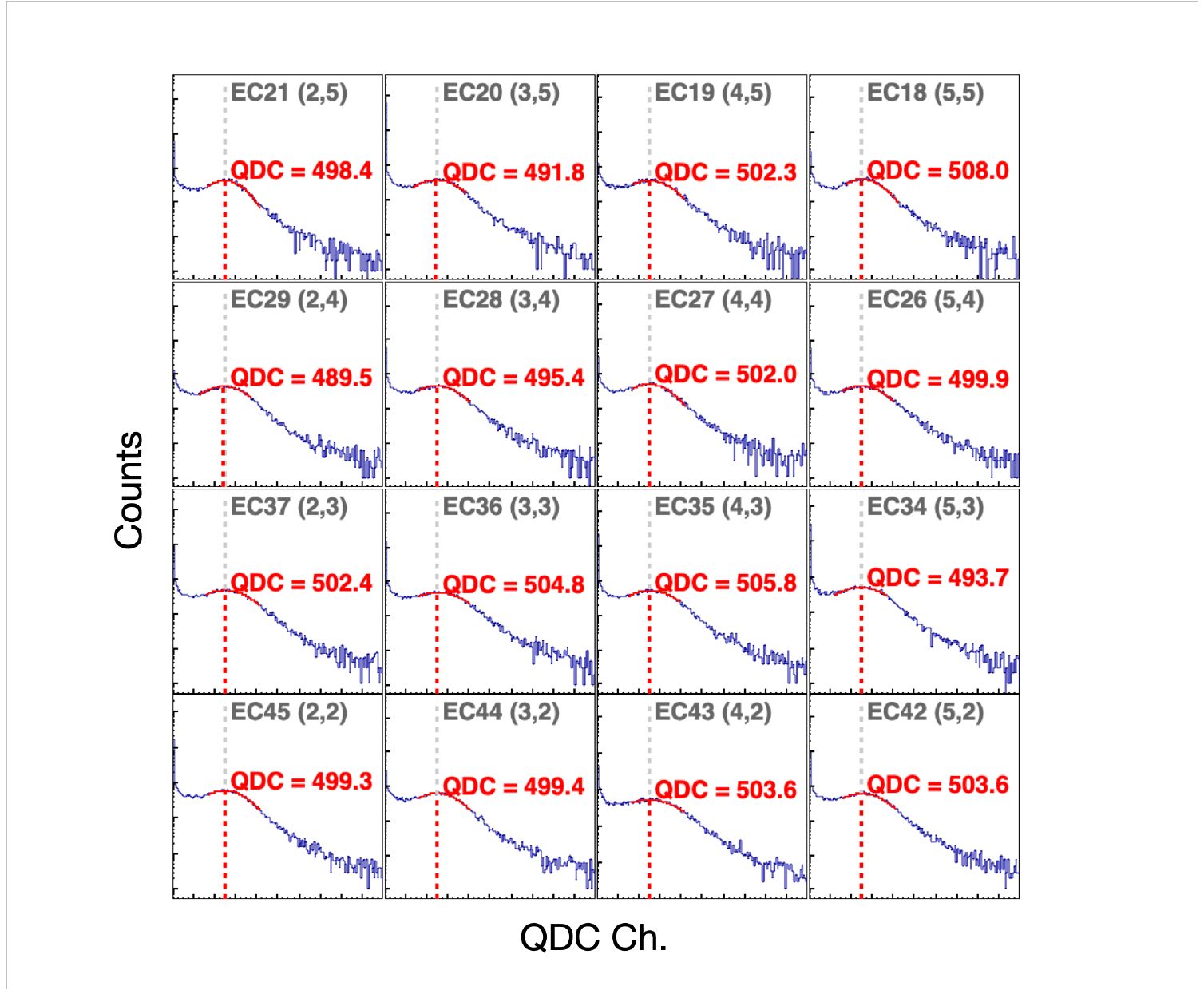}
\caption{\label{fig:gain_match_bar} Semi-log graphs of QDC spectra of the central 16 crystals after gain matching using cosmics. QDC peaks are aligned to channel 500 with software calibrations.} 
\end{figure}

%The QDC gain match of the calorimeter is done using cosmic rays. 
The gain is calibrated in two steps.
First, a hardware gain match adjusts PMT HV.
Second, the QDC response of each channel is fine tuned in software. 
During the hardware gain matching, 
%two scintillator bars  placed below the calorimeter are used to trigger the system for cosmic ray particles. 
we adjust PMT high voltages to approximately match the positions of the peaks in the QDC spectra of each crystal.
%in the calorimeter are compared.
The Hamamatsu R1355 PMT has 10 dynode stages. 
The relationship between applied high voltage and gain was found to be
\begin{equation}
\frac{\rm{gain}_{new}}{\rm{gain}_{old}} = \left(\frac{V_{\rm{old}}}{V_{\rm{new}}}\right)^k ,
\end{equation}
where $k$ was determined empirically to be approximately 8, with some variation between crystals. 
Following this relation, the high voltage applied to each channel of the detector is adjusted iteratively until the QDC peak positions of the cosmic events for each crystal are roughly matched at around QDC channel 500. 
This setting 
%maximizes the gain 
optimizes the calorimeter performance while keeping the QDC spectra from overflowing at higher beam momentum settings.
At the highest beam momentum setting for data, 210 MeV/$c$, the QDC peaks are at about channel 3100, with a long tail reaching to channel 3840, near the end of the 12-bit QDC range.
After the hardware gain match, in the software gain matching step, we fine tune the gain match by fitting the QDC peak of the cosmic signal for each bar, determining the factor needed to scale the peak position to the common QDC channel of 500.
%the QDC peak of the cosmic signal for each bar is fit with a Gaussian to find the peak position.
%mean of the distribution. 
%This determines a factor needed to fine tune the peak positions to the common QDC channel of 500.
%By calibrating the cosmic QDC peak to channel 500, the QDC peaks at about 2100 with a long tail speading to ch 3840, which is the highest QDC ch we read out, at the highest momentum setting for production data taking.
An example of the QDC spectra of the central crystals after the gain matching calibration is shown in Fig.~\ref{fig:gain_match_bar}. 

% \begin{figure}[h]
% \centering
% \includegraphics[width=0.45\textwidth]{gain_match_quality.png}
% \caption{\label{fig:gain_match_check} 
% (Top) Data for position of QDC peak in units of QDC channels vs.\ the number of the calorimeter crystal with the maximum energy, and the percent difference of that peak position from the average of the 9 bar sums. The crystal numbers increment across the rows of the calorimeter crystals, from the top row to the bottom.
% We perform 9-bar sums for central crystals, 6-bar sums for side crystals, and 4-bar sums for corner crystals.
% (Bottom) Same as top, but for the Geant4 simulation.
% %from sum average of up to 9 adjacent bars vs. bar number.
% } 
% \end{figure}

\begin{figure} [h]
\centering
\includegraphics[width=0.49\textwidth]{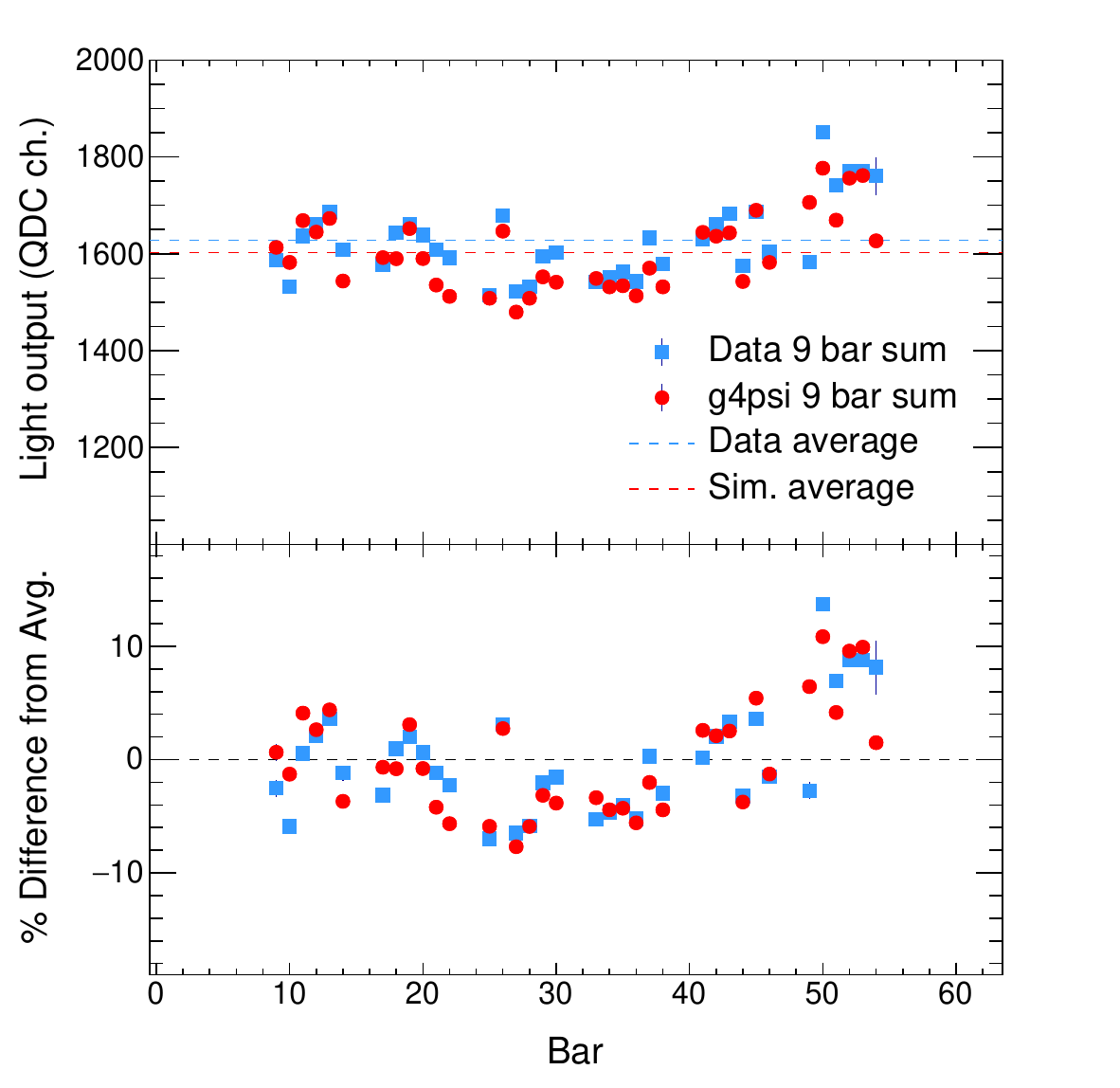}
\caption{\label{fig:gain_match_check} 
Data for position of the calorimeter QDC peak in units of QDC channels vs. the channel number of the crystal with the maximum light output, and the percent difference of that peak position from the average of the 9 bar sums, at 110~MeV/$c$. The crystal numbers increment across the rows of the calorimeter crystals, from the top row to the bottom.} 
\end{figure}

Finally, to check the uniformity of the gain matching, the cluster energies are determined as a function of position in the calorimeter.
%from the detector for clusters at different locations are compared to see if the calibration is uniform across the detector.
Figure~\ref{fig:gain_match_check} shows the result of this study, where the cluster light output sum vs. central bar location for 110 MeV/c beam data are plotted. 
From the figures, the average light output of clusters with a 9-bar sum is roughly the same independent of their position. %with some decrease in the energy for crystals at the edge of the calorimeter..
To better understand the distribution, we compare the data to simulation.
% To better understand the distribution, the detector response to 1-GeV muon showers was simulated.
%In this simulation, a test plane with area and position corresponding to the cosmic trigger scintillators is used to trigger the system. 
The simulation reproduces the behavior of the light output peak vs.\ crystal position with maximum light output.
% The main observation is that edge crystals give slightly smaller energy sums compared to central crystals.
% Also, upper crystals of the detector (smaller number bars)  give slightly smaller energy sums compared to the bottom crystals (bigger bar numbers). 
More details about the calorimeter simulation and data digitization will be discussed in later sections.

%{\it\color{red} Perhaps the top and bottom blocks have lower gain for cosmics as showers take several cm to develop, so cosmics first entering these blocks generate less light. But how to explain the bottom blocks?}

\section{\label{sec:performance}Detector Performance}

For each event, the sum of light output in the calorimeter is determined.
First, the crystal with the highest QDC value is found. 
Second, the light output of that crystal and the eight surrounding neighbors are summed. 
%{\it\color{red} Is the TDC time of this first cluster checked for being in time?}
Third, a search is performed for additional  clusters away from the first cluster found.
TDC information is checked for each cluster to determine if it is in-time, so that energy cuts should be applied.

\begin{figure}[h]
\centering
\includegraphics[width=0.49\textwidth]{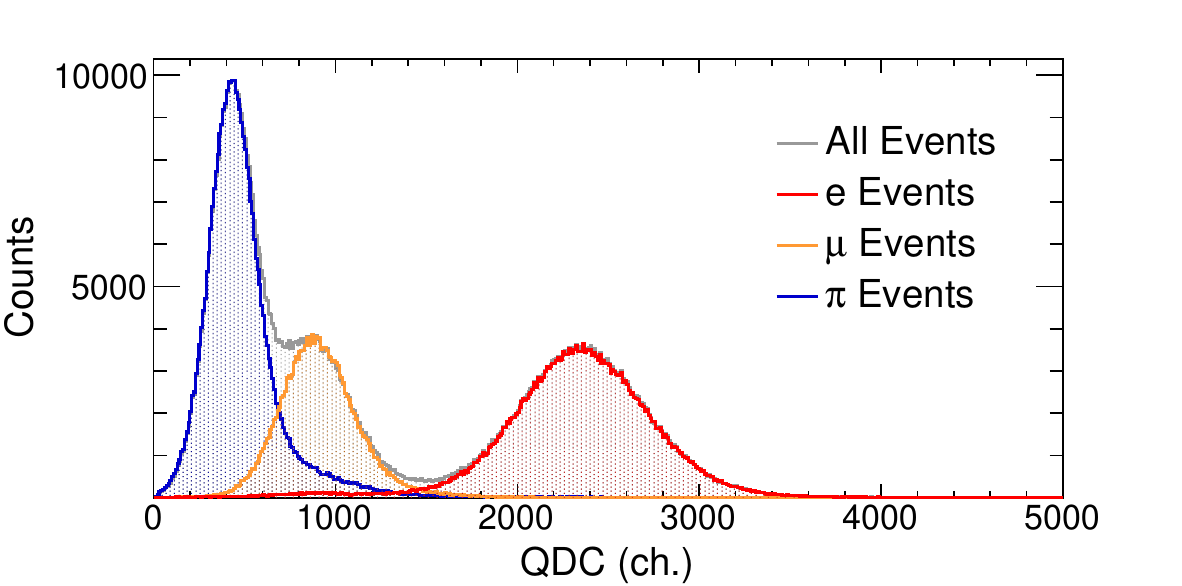}
\includegraphics[width=0.49\textwidth]{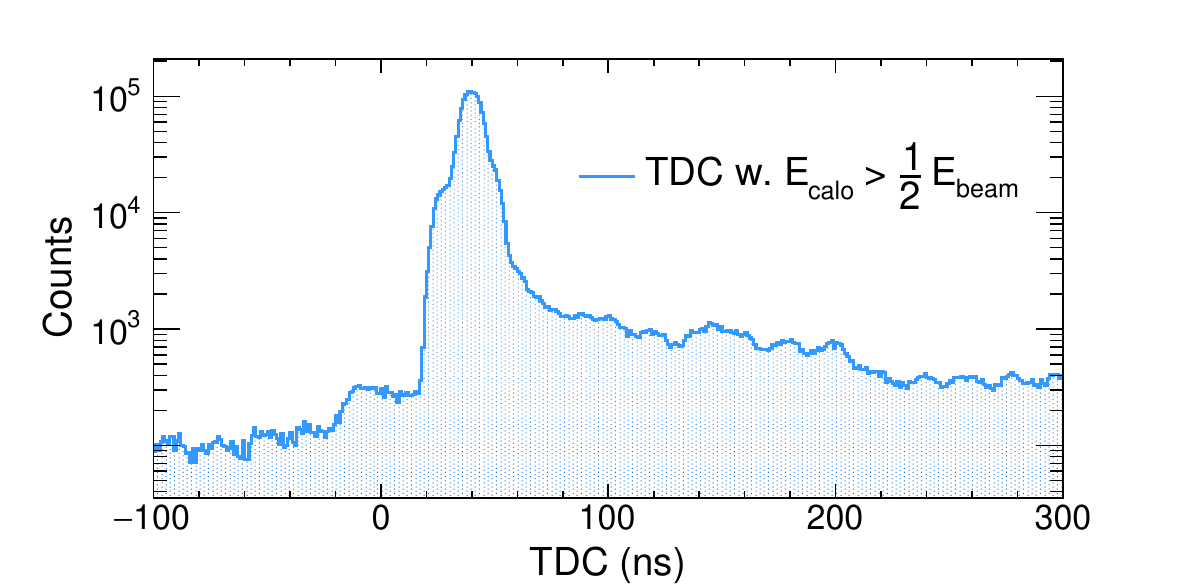}
\caption{\label{fig:calo_sig_raw} \textbf{Top:} Distribution of calorimeter light output 9-bar sums for beam particle events taken at momentum of 160~MeV/$c$. 
The distributions from different particle types are normalized to the same peak height as the distribution for all events.
\textbf{Bottom:} TDC times relative to the event trigger for events with light output greater than half of the beam energy.
The in-time peak is at $\approx$ 40 ns.
%due to an arbitrary time offset.
} 
\end{figure}

Figure~\ref{fig:calo_sig_raw} shows the 9-bar QDC sum for the calorimeter at 160 MeV/$c$ beam momentum.
Data were taken with a beam particle trigger at approximately 180 kHz beam rate,
leading to an accidental coincidence rate within the 200-ns QDC gate of approximately 4\%.
Incident particle species are identified using timing information from the BH with respect to the accelerator RF \cite{my_thesis}. 
%In Fig ~\ref{fig:calo_sig_raw}, the QDC sum of the different particle types are also shown.
The bottom of Fig.~\ref{fig:calo_sig_raw} shows the TDC distribution of the calorimeter for events with light output greater than half of the beam energy. 
The in-time peak at approximately 40 ns is for particles triggering the system. 
The small background at earlier times reflects particles that did not trigger the DAQ system, as it was still processing a previous event.
%It is less likely to have events before the trigger peak, because earlier events tend to assign as the trigger.
In contrast, particles arriving after the triggering particle will generate signals, unless they overlap the triggering particle by being in the same calorimeter crystals within the 25-ns TDC dead time. 
%peak within the trigger window will be recorded. 
Hence, the random coincident background is higher after the trigger peak than before the trigger peak. 
The 19.75-ns beam RF structure leads to time structure in the random background events.
It is washed out by the multiple particle types at different phases of the RF convoluted with 10-ns FPGA clocks in the trigger.

To distinguish the photon events from the leptons and pions, the BM detector in front of the calorimeter is used. 
Photons are not expected to leave a signal in the thin scintillators of the BM, while electrons, muons and pions will deposit energy in the detector. 
%To distinguish the different particle types from the mixed particle beam, we use the beam hodoscope timing with respect to the accelerator RF for particle identification.
In the following sections, the calorimeter detector performance will be shown using electrons, which generate similar signals in the calorimeter to photons of the same energy. 
%{\it\color{red} I believe this, but have we checked it with Geant4?}
%Since the PiM1 beam is a mixture of electrons, muons and pions, particle identification is necessary and is done by the timing from the Beam Hodoscope. 

\subsection{Energy Response} \label{subsec:energy_response}

\begin{figure} %[h]
    \includegraphics[width=0.45\textwidth]{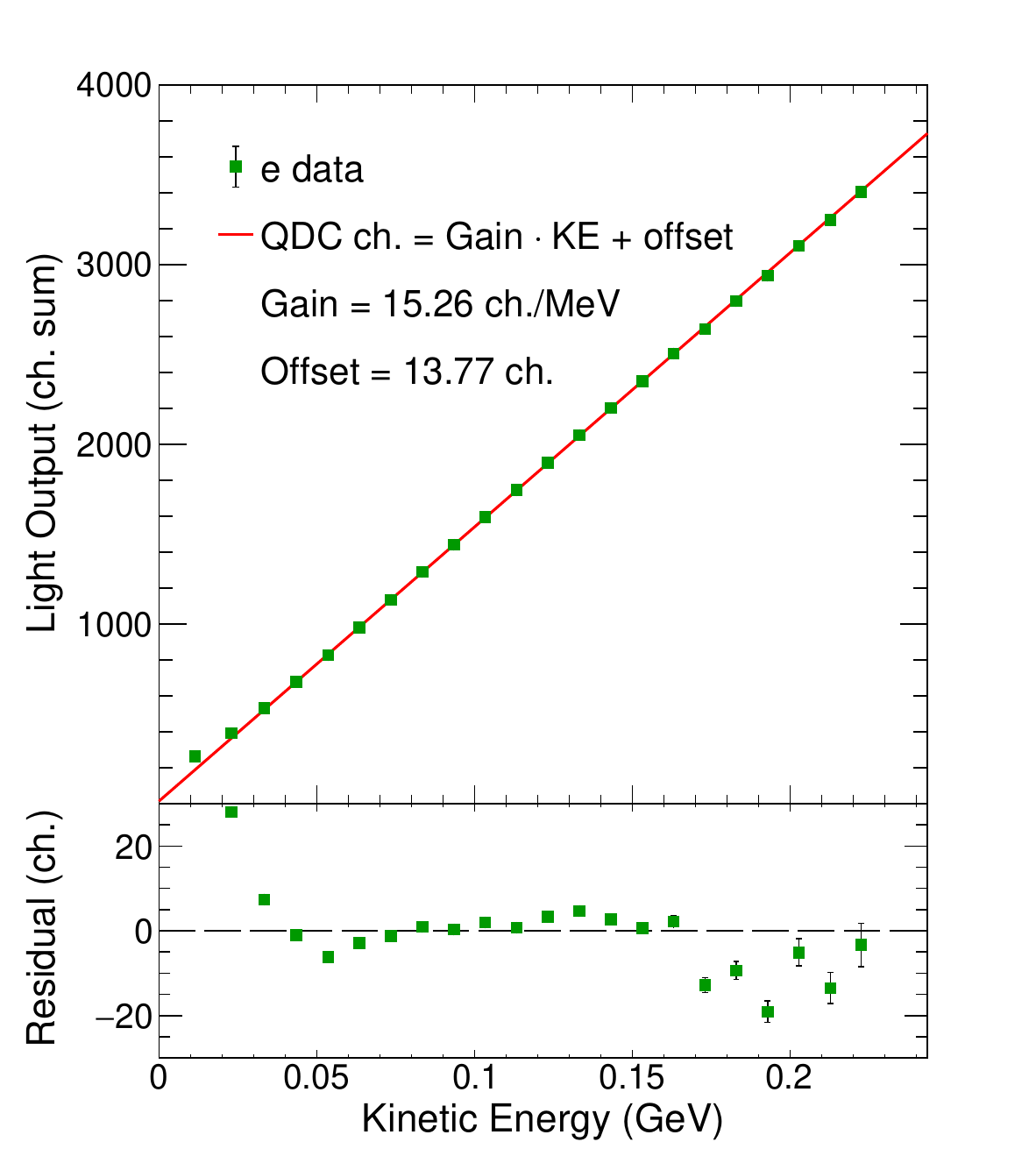}
    \includegraphics[width=0.45\textwidth]{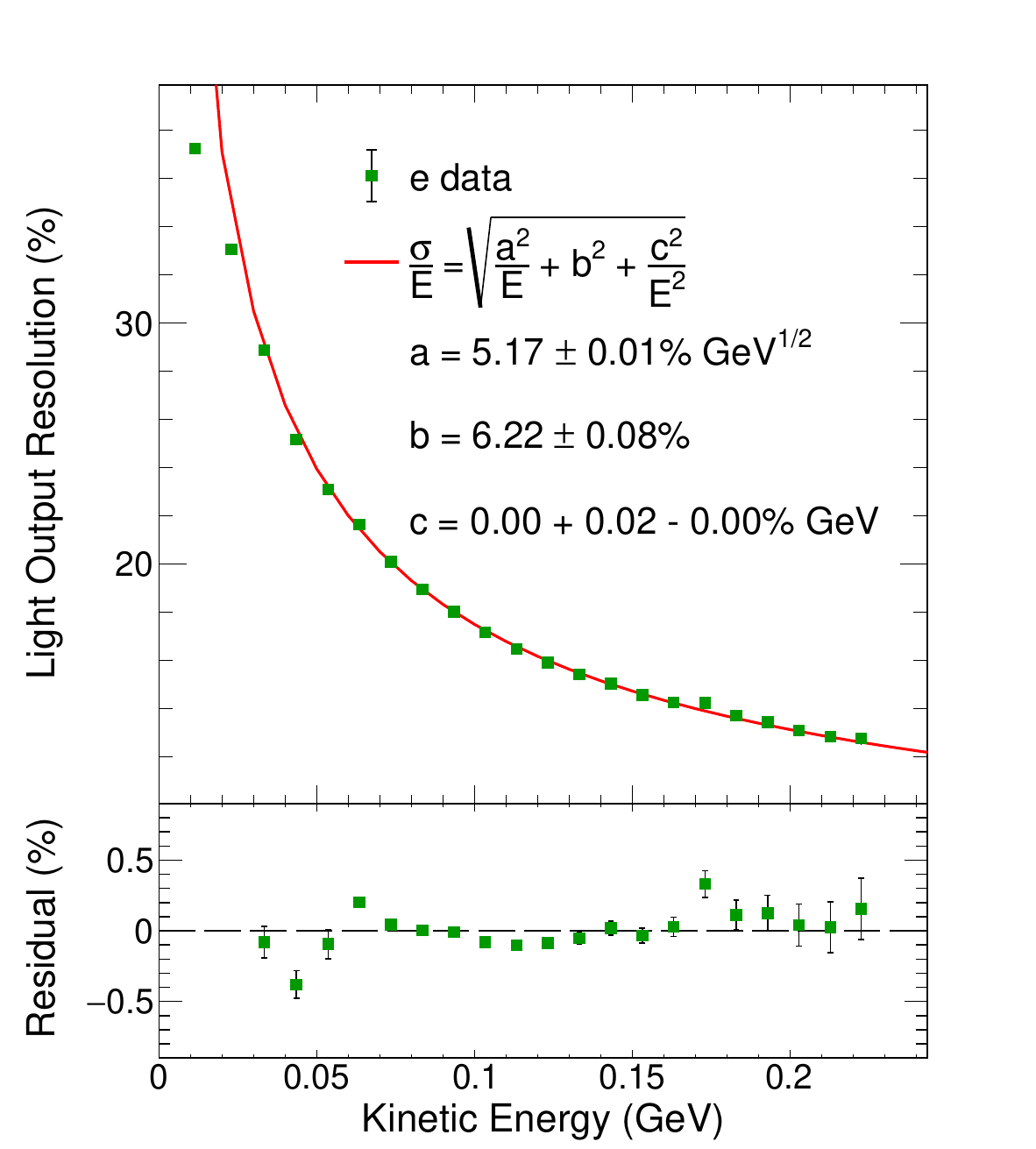}
    \caption{\label{fig:energy_response} Energy scan results for positrons. \textbf{Top:} Light output vs. beam momentum. \textbf{Bottom:} Resolution vs. beam momentum. Residuals of the fits are shown at the bottom of each plot with at most two energy outliers off the scale.} 
\end{figure}

Beam data were used to study and calibrate the calorimeter absolute light output response as shown in Fig.~\ref{fig:energy_response}.
The PiM1 beam line is designed to operate at momenta above about 100 MeV/$c$~\cite{beam_paper}.
To map out the detector response in a wide range of energy, data were taken at beam momenta from 20 MeV/$c$ to 230 MeV/$c$, and simulations were used to estimate the energy of the particles entering the calorimeter.
% Data were taken at beam momenta %settings 
% from 20 MeV/$c$ to 230 MeV/$c$ to study and calibrate the calorimeter absolute energy response.
%{\it\color{red} 
%Do we need a specification for how well we know the incoming beam momentum?
%Need to check how much energy is lost from the channel to the calorimeter for mu's and pi's.
%}
% Figure~\ref{fig:energy_response} shows the results of this momentum scan. 
The momentum scan result shows that the light output in the calorimeter as a function of the beam energy is very close to linear.  
%\textcolor{red}{distribution}.
%\textcolor{red}{A linear function is fit to the data} 
The data are fit with a linear function to check linearity and to determine parameters to be used in the simulation to model the detector light output. 
The light output resolution of the detector is fit with 
%a square root function,
$\frac{\sigma}{E} = \sqrt{\frac{a^2}{E}+b^2+\frac{c^2}{E^2}}$, which is commonly used to describe the resolution of calorimeters~\cite{pdg_calo}. 
In this fit, $a$ is the stochastic term that is governed by the electromagnetic shower fluctuations in the material, $b$ is the systematic term that reflects the uniformity of the detector and how well the detector is calibrated, and
$c$ is the noise term from electronic readout when measuring the energies.
Typical light output resolutions for lead glass calorimeters are about $5\%/\sqrt{E/\rm{GeV}}$ \cite{pdg_calo}. 
For the MUSE calorimeter, the parameters from the fit in Fig.~\ref{fig:energy_response} indicates that the resolution from the stochastic term is about $5.17\%/\sqrt{E/\rm{GeV}}$. About $6.22\%$ resolution arises from the $b$ calibration term, while less than $1\%$ is from the electronics.

\subsection{Detector Timing}

\begin{figure} [h]
    \centering
    \includegraphics[width=0.49\textwidth]{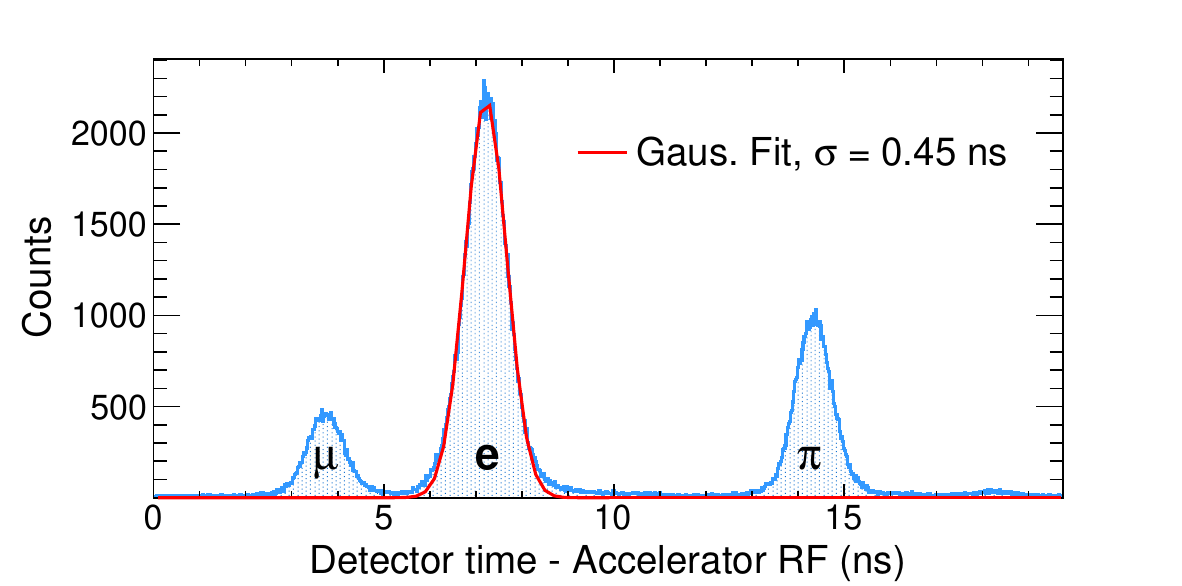}
    \includegraphics[width=0.49\textwidth]{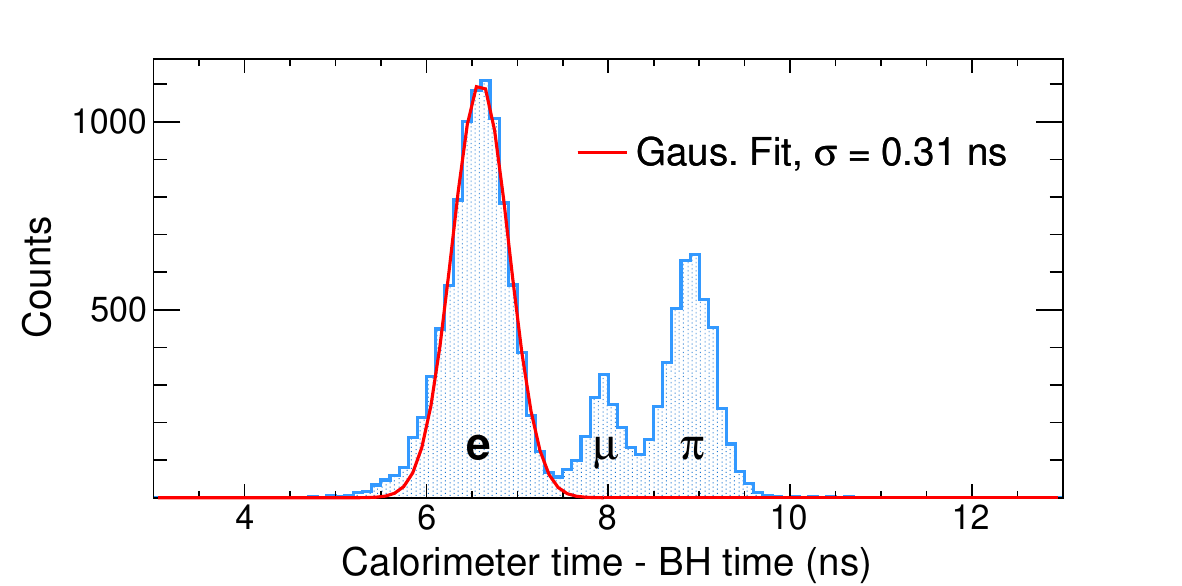}
    \caption{TDC times of a central calorimeter crystal at 160 MeV/$c$, before walk corrections.
    \textbf{Top:} RF time, the difference in time between the calorimeter signal and the accelerator RF signal.
    \textbf{Bottom:} Time difference between calorimeter and BH signals, offset by an arbitrary constant.
    Walk corrections have not been applied to the calorimeter timing.
    }
    \label{fig:calo_timing}
\end{figure}

Light output in the calorimeter from randomly coincident beam particles can be identified by two techniques.
The BM hodoscope immediately upstream of the calorimeter localizes incoming charged particles at the cm level in position and at the 100 ps level in time.
The light output from these beam particles is known from the beam momentum.
The MUSE calorimeter also reads out timing information for the calorimeter crystals to help distinguish multiple hits in the detector, from both charged particles and photons. 
Figure~\ref{fig:calo_timing} shows the example of calorimeter timing for a central crystal at 160 MeV/$c$. 
The top plot shows the detector time relative to the accelerator RF (modulo the RF period). 
The distribution shows three distinct peaks for the three particle types in the beam. 
The bottom plot shows time of flight from the beam hodoscope to the calorimeter, over a flight path of approximately 2.2 m.
The electron, muon and pion peaks are also observed. 
%The timing resolution of both plots is less than 0.5 ns. 
%Since the accelerator has a RF period of 19.75 ns, the timing resolution of the calorimeter is sufficient in identifying the random coincident particles from the beam. 
The separation of the timing peaks of different particles shown in Fig.~\ref{fig:calo_timing} makes clear that the timing information is sufficient to detect, localize and identify randomly coincident beam particles, other than those in the same crystals within the electronic dead times, approximately 25 ns.

%There are two significant factors in our treatment here that degrade the timing resolution of the calorimeter compared to other timing detectors in the MUSE system.
%First, dispersion from the delay of the calorimeter signal before the MCFD slows the rise time and reduces the resolution.
%This was needed due to technical limitations in our system, and cannot be easily improved.
%Second, the particles leave a wide range of energy in the calorimeter, as can be seen from the resolutions in Fig.~\ref{fig:energy_response}.
%Walk corrections have not been implemented for the calorimeter timing, as, while they are desirable, they are not needed for our analysis.
%This effect on the timing is most evident in Fig.~\ref{fig:calo_timing}, where the $\mu$-$e$ ($\pi$-$e$) time of flight difference is $\approx$ 0.1 (0.2) ns longer than the expected 1.3 (2.2) ns, due to the smaller signals and consequent later timing of the more massive particle.
% for example, electron QDC spreads out over 2000 channelsThe example plots have not included walk corrections. There are two significant factors that contribute to the resolution. . The resulting resolution of the calorimeter timing is much larger than other scintillator detectors such as the BH and SPS. Time walk correction will be done to improve the timing resolution despite the fact that the current timing resolution already meets the requirement of the detector for the experiment.

\subsection{Simulation}

\begin{figure}[h]
\centering
\includegraphics[width=0.45\textwidth]{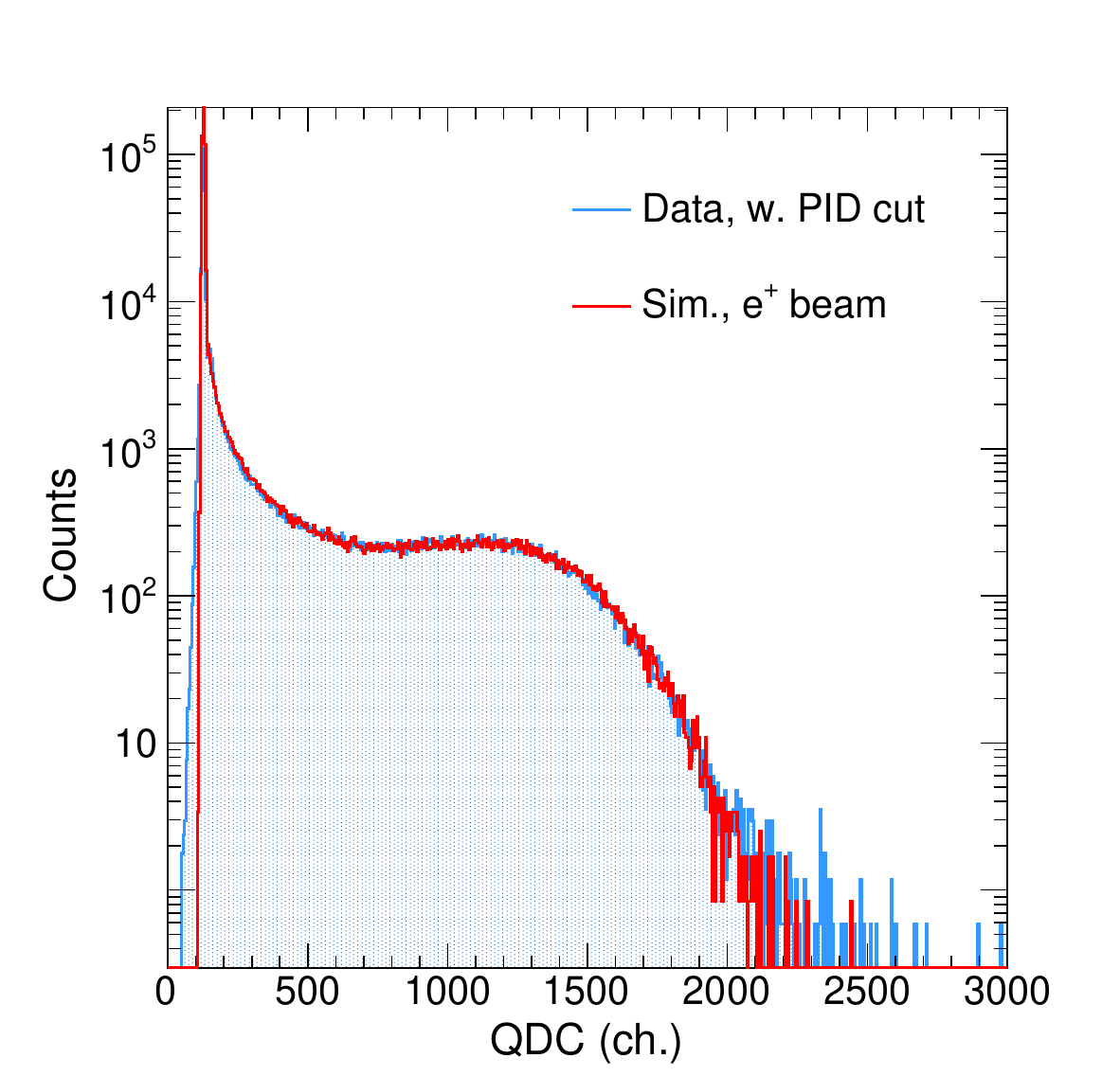}
\includegraphics[width=0.45\textwidth]{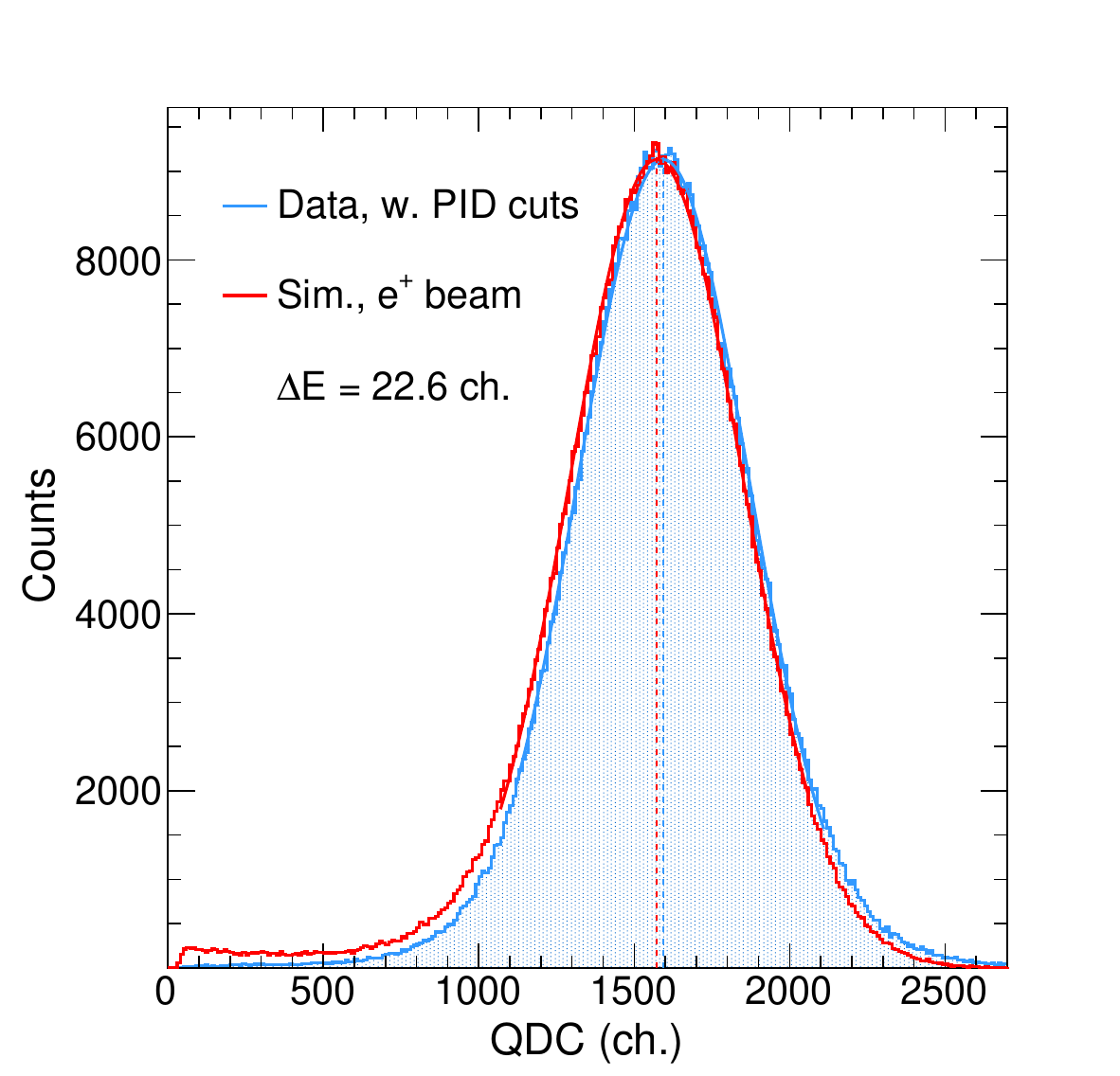}
\caption{\label{fig:calo_sim} \textbf{Top:} Positron QDC spectra of data and simulation for a central crystal. \textbf{Bottom:} Positron light output 9-bar sum for data and simulation at +110 MeV/c.} 
\end{figure}

The MUSE setup is simulated with Geant4. 
The simulation records hits and saves the events to a ROOT output tree. 
The simulated data is digitized and run through the same analysis as experimental data to allow a direct comparison of simulation to data. 
For the calorimeter, the simulation records Cherenkov light production in the crystals.
There are two modes of simulation that can be selected: a fast simulation that integrates a light yield over the path in the crystal for $\beta > c/n$, and a more detailed simulation that creates optical photons at each reaction step and tracks those that reach the PMTs.
Current indications are that it is not necessary to run the simulation in the much more computationally intensive photon-counting mode. 
Here, we will show the preliminary calorimeter simulation result using the fast simulation mode.
In this mode of the simulation, we calculate the light output by estimating the number of photons produced by the Cherenkov radiation.
Since the number of photons emitted by incident particles is proportional to $(1-\frac{1}{n^2\beta^2})$, for each reaction step in the simulation, we can find the relation for the number of photo-electrons read out in an event by the calorimeter as 
\begin{equation}
    \delta N = f \times \delta l \times \left(1-\frac{1}{n^2\beta^2}\right),
\end{equation}
where $\delta l$ is the step length, $n$ is the index of refraction, and $f$ is scaling factor tuned to %roughly 
match the simulation to data.
Integrating $\delta N$ for all the steps in the crystal determines the total number of photo-electrons generated by particles 
%energy deposited 
in the crystal. %{\color{red}Note: We need to rephrase the last paragraph in terms of number of photons and light output, not energy deposition, $\delta E \to \delta N$, etc.}

%Another mode of the simulation by counting the photons generated by the scintillator and collected by the PMTs can be activated for more realistic simulation. However, the running speed of the photon counting mode is much slower than the Cherenkov light mode. For the following discussion, results from the fast Cherenkov light mode are shown. 

%While the simulation records the energy deposited in the calorimeter, t
The digitization converts the simulated light output of the calorimeter crystal to detector QDC channels. 
The pedestal of each QDC spectrum is modeled with a Gaussian distribution with the same width as the data, which is found during the calibration. 
If the bar has energy deposited in it, the energy is multiplied by a factor to more precisely model the gain of the detector and then randomized with a Gaussian, where the width of the distribution is equal to 
\begin{equation}
    \sigma=E\sqrt{\frac{\alpha^2}{E}+\frac{\gamma^2}{E^2}}.
\end{equation} 
The values of parameter $\alpha$ ($\approx4.46\%~\rm{GeV}^{1/2}$) and $\gamma$ ($\approx0.066\%~\rm{GeV}$) are similar to the parameter $a$ and $c$ in the resolution fit of the light output scan calibration shown in Fig.~\ref{fig:energy_response}. Because the simulation has some resolution effects already when calculating the light output for each reaction steps in the crystal, the additional resolution needed at the digitization is smaller than the fit values from the data.
% The three parameters in this equation along with the gain parameters are tuned until the QDC spectrum of each bar obtained from simulation matches the data.
%Figure~\ref{fig:calo_sim}  compares the QDC spectra of data and simulation for a central bar of the calorimeter, tuned at 110 MeV/c. 

% \begin{figure}[h]
% \centering
% \includegraphics[width=0.45\textwidth]{sim_calib_ch.pdf}
% \caption{\label{fig:calo_sim_ch}  QDC spectra of data and simulation for a central crystal.} 
% \end{figure}

% \begin{figure}[h]
% \centering
% \includegraphics[width=0.45\textwidth]{sim_sum.pdf}
% \caption{\label{fig:sim_sum} Energy sum of data and simulation at 110 MeV/c.} 
% \end{figure}

\begin{figure} [h]
\centering
\includegraphics[width=0.45\textwidth]{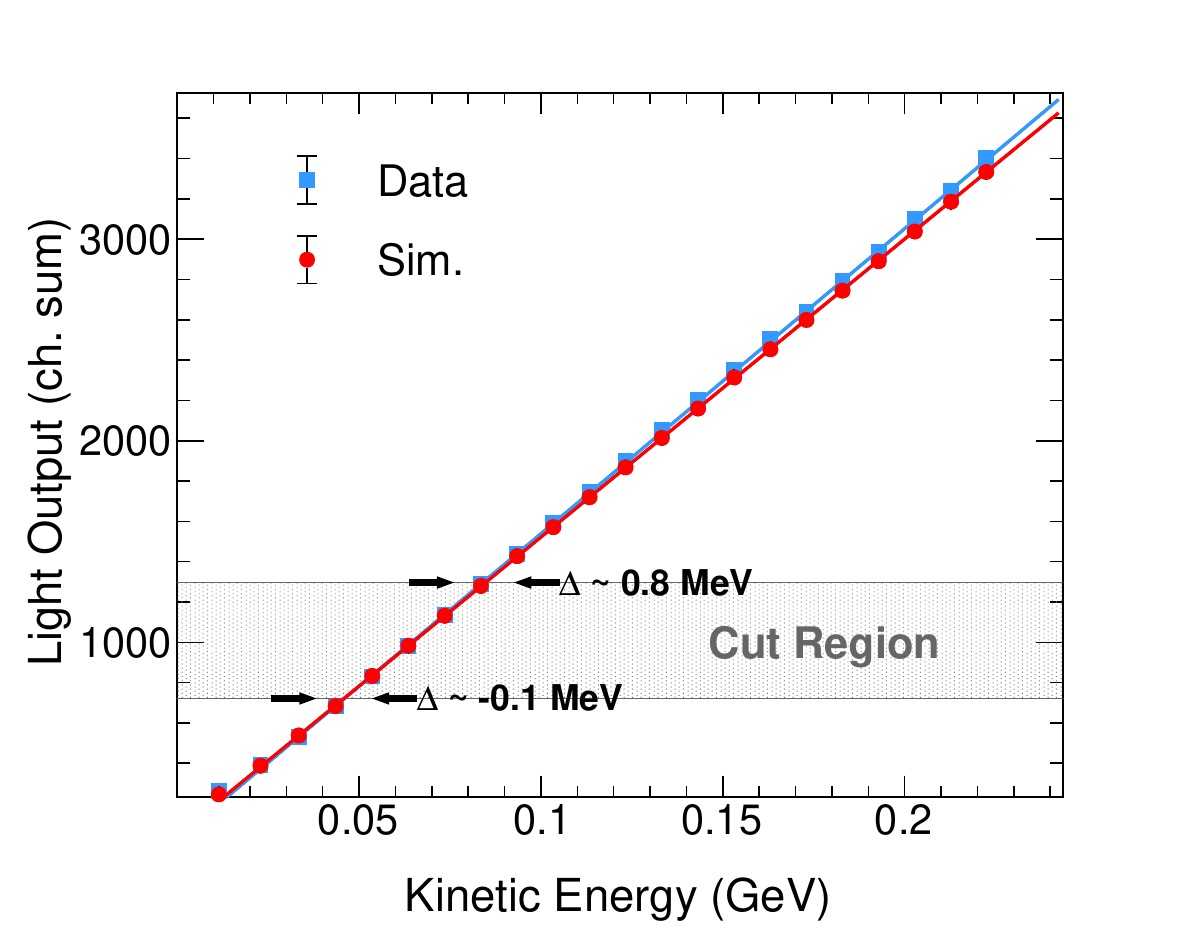}
\caption{\label{fig:sim_light_output} Comparison between data and simulation for calorimeter light output sum (in units of QDC channels) for electrons vs.\ beam momentum. 
Linear fits are performed to compare differences at different beam energies.
The arrows show the differences between data and simulation for the QDC values for the default radiative corrections cuts for 115~MeV/$c$ (bottom of the gray band) and 210 MeV/$c$ (top of the gray band).}
\end{figure}

\begin{figure} [h]
\centering
\includegraphics[width=0.45\textwidth]{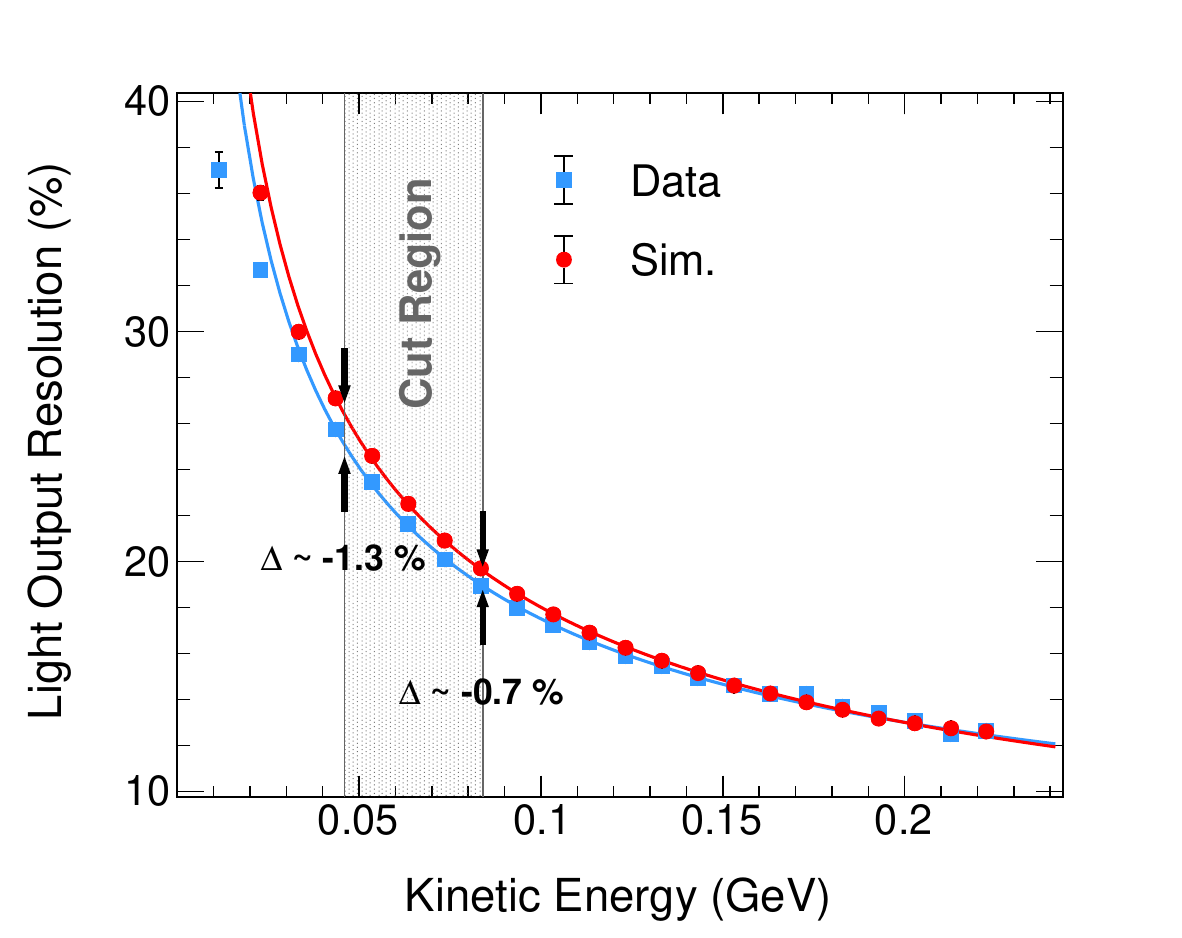}
\caption{\label{fig:sim_resolution} Comparison between data and simulation for calorimeter light output resolution for electrons vs.\ beam momentum. 
Resolution fits as discussed in Sec.~\ref{subsec:energy_response} are performed to compare differences at different beam energies.
The arrows show the differences between data and simulation for the default radiative corrections cuts for 115 MeV/$c$ (left edge of the gray band) and 210~MeV/$c$ (right edge of the gray band).
%Differences between data and simulation at where we will apply cuts for 115 MeV/c and 210 MeV/c data are shown by the arrows.
} 
\end{figure}

%After tuning, 
The digitized, simulated data are processed through the same analysis as the experimental data.
%Here we compare the simulations and measurements.
Figure~\ref{fig:calo_sim} compares data and simulation for the spectrum of a single crystal and for the 9-bar light output sum from the calorimeter.
%energy sum comparison between data and simulation. 
The data and simulation agree well, with a small mismatch at the mean value. 
This difference is due to a mismatch in the energy to QDC conversion between data and simulation for some channels, along with small differences in the air gap between bars. 

%Figure~\ref{fig:calo_sim_vs_data} shows the energy response and energy resolution from data and from simulation at different momentum settings. 
Figures~\ref{fig:sim_light_output} and~\ref{fig:sim_resolution} compare the 9-bar-sum light output response and resolution of data and simulation at different momentum settings. 
Both data and simulation show similar linear relationships in the light output response, and similar light output resolution. 
While there is some disagreement at higher momenta, in the region where event cuts will be applied for radiative corrections (40\% of the beam energy), the differences are small and the agreement is better than our 2-MeV requirement.
%\textit{\color{red} I don't think specification was actually given before.}
Note that the energy to QDC channel conversion presented here is based on a calibration at one momentum setting, 110 MeV/$c$ --
no tuning was done to adjust the light output response of the simulation to match the data. 
%, and the simulation is still at a preliminary stage. Nevertheless, simulation and data agree well and show promising results.

\section{\label{sec:Photons}Reconstructed Photons}

\begin{figure} [h]
\centering
\includegraphics[width=0.49\textwidth]{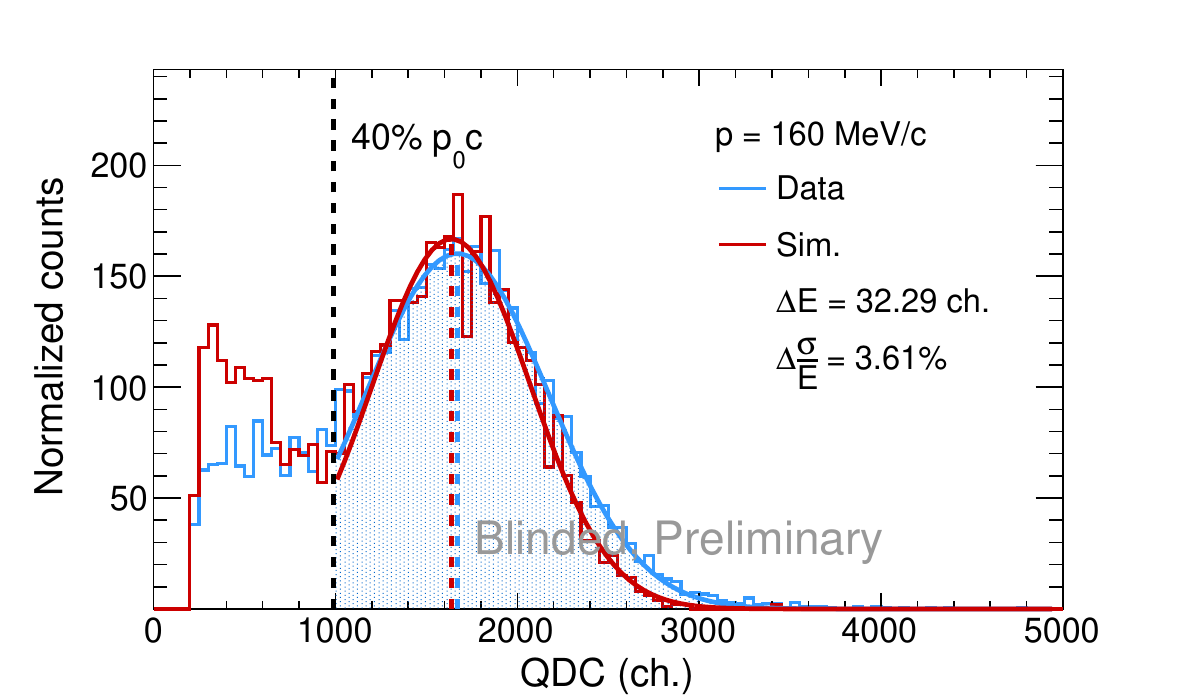}
\caption{\label{fig:Rec_Photon} Reconstructed photon QDC spectra of data and simulation for 160 MeV/c electron scattering with liquid hydrogen target. Black dashed line on the left indicates the default radiative correction cut for this momentum setting. Blue and red dashed lines indicate the mean position of the Gaussian fit of the high energy photon peaks.}
\end{figure}

Figure~\ref{fig:Rec_Photon} presents an example of the light output distribution of the reconstructed photons in the calorimeter for data and preliminary simulation. 
The default radiative correction cut at $E_{\gamma} < 0.4 p_{0}c$, where $p_0$ is the beam electron momentum, is indicated by the dashed line. 
The two distributions are similar in shape for the high energy photon events, with simulation being slightly narrower in width.
The slight mismatch in resolution is expected as the simulation is at a preliminary stage with the calorimeter energy calibration based on one momentum setting only, and as the scattered-particle scintillator energy cuts have not been precisely matched in simulation and in data. 
The difference in the lower energy events is due to the difference in threshold setting between data and simulation. 
%Nevertheless, b
Both data and simulation show a prominent peak from the high-energy photons emitted by some scattering electrons, as expected. 
The radiative correction cut will remove these events, reducing the experiment's sensitivity to radiative corrections. 
The agreement between the calibrated data and simulation indicates that the calorimeter performance is sufficient to obtain the needed experimental uncertainties. 
Further tuning in the simulation calibration, including calibration using data taken at multiple energies and studying possible time dependence of the calorimeter response will improve the agreement to be better than requirements.
%data and simulation will reach better agreement. 

\section{\label{sec:summary}Summary}

In order to have radiative corrections under control for the MUSE scattering experiment, a lead-glass calorimeter detector was built to capture high-energy photons from initial-state radiation. 
%The detector is sufficient and is large enough to have a good coverage on the forward going photons. 
Studies show that the detector has an light output response and resolution sufficient to identify and remove these events.
The understanding of the detector behavior is demonstrated by comparing calibration measurements to simulation. 
With the calorimeter,  MUSE  will be able to test and control radiative corrections.

% \textit{\color{red} Revised opinion... I think it would be good to have a spectrum of calo energy for scattered electron events, with single in-time BH and no in-time BM cuts, comparing data and simulation. Forget the others I had mentioned. Not needed for this paper.
%\begin{itemize}
%    \item beam 
%    \item beam electrons
%    \item scattered particles
%    \item scattered electrons
%    \item I wonder if these should be shown with wide cuts to ensure energy from only one particle vs without such cuts
%    \item and at some point we need to evaluate how well spectra without such wide cuts on multiple beam particles but with calo time and one cluster cuts reproduce the wide time cut spectra 
%\end{itemize}.ra
% }

\section{\label{sec:acknowledgements}Acknowledgements}

This work is supported by the National Science Foundation, grants NSF PHY-1913653, PHY-2209348, PHY-2110229, PHY-2111050, PHY-2310026, PHY-2113436, PHY-1812402, PHY-1505934, PHY-2012114,  PHY-2012940, PHY-2310026, PHY-10011349, and HRD-1649909, the U.S. Department of Energy Office of Science, office of Nuclear Physics under contract nos. DE-SC0016577, DE-SC0019768, DE-AC02-06CH11357, DE-SC0012589, and DE-SC0014448, the US-Israel Binational Science Foundation (BSF) Grant 2020796, Schweizerischer Nationalfonds (SNF) 200020-156983, 132799, 121781, 117601, and the Paul Scherrer Institute. M.~Kohl has been jointly supported by Jefferson Lab.

We are deeply grateful to the Mainz A2 collaboration for lending us the lead-glass crystals, which were essential in constructing the calorimeter. We thank the Paul Scherrer Institut for providing the necessary facilities, beam and resources for our research. We acknowledge PSI ``Hallendienst" for their help in reinforcing the MUSE platform for the calorimeter.

\bibliography{apssamp}% Produces the bibliography via BibTeX.

\end{document}

%% file: authorlist.tex
\author{W.~Lin}
\altaffiliation{Current: Center for Frontiers in Nuclear Science \& Department of Physics and Astronomy, Stony Brook University, Stony Brook, NY 11794, USA}
\affiliation{Department of Physics and Astronomy, Rutgers, The State University of New Jersey, Piscataway, NJ 08855, USA}

\author{T.~Rostomyan}
\affiliation{Laboratory for Particle Physics, PSI Center for Neutron and Muon Sciences, 5232 Villigen, CH}
% \affiliation{Laboratory for Particle Physics, PSI Center for Neutron and Muon Sciences, Forschungsstrasse 111, 5232 Villigen PSI, CH}

\author{R.~Gilman}
\affiliation{Department of Physics and Astronomy, Rutgers, The State University of New Jersey, Piscataway, NJ 08855, USA}

\author{S.~Strauch}
\affiliation{Department of Physics and Astronomy, University of South Carolina, Columbia, SC 29208, USA}

\author{C.~Meier}
\affiliation{Department of Physics, University of Basel, 4056 Basel, CH}

\author{C.~Nestler}
\affiliation{Laboratory for Particle Physics, PSI Center for Neutron and Muon Sciences, 5232 Villigen, CH}
% \affiliation{Laboratory for Particle Physics, PSI Center for Neutron and Muon Sciences, Forschungsstrasse 111, 5232 Villigen PSI, CH}
\affiliation{Institute for Particle Physics and Astrophysics, ETH Zürich, 8092 Zürich, CH}

\author{M.~Ali}
\affiliation{Department of Physics, New Mexico State University, Las Cruces, NM 88003, USA}

\author{H.~Atac}
\affiliation{Department of Physics, Temple University, Philadelphia, PA 19122, USA}

\author{J.~C.~Bernauer}
\affiliation{Center for Frontiers in Nuclear Science, Stony Brook University, Stony Brook, NY 11794, USA}
\affiliation{Department of Physics and Astronomy, Stony Brook University, Stony Brook, NY 11794, USA}

\author{W.~J.~Briscoe}
\affiliation{Department of Physics, The George Washington University, Washington, DC 20052, USA}

\author{A.~Christopher Ndukwe}
\affiliation{Department of Physics, Hampton University, Hampton, VA 23668, USA}

\author{E.~W.~Cline}
\affiliation{Center for Frontiers in Nuclear Science, Stony Brook University, Stony Brook, NY 11794, USA}
\affiliation{Laboratory for Nuclear Science, Massachusetts Institute of Technology, Cambridge, MA 02139, USA}

\author{K.~Deiters}
\affiliation{Laboratory for Particle Physics, PSI Center for Neutron and Muon Sciences, 5232 Villigen, CH}
% \affiliation{Paul Scherrer Institute, Villigen 5232, CH}

\author{S.~Dogra}
\affiliation{Department of Physics and Astronomy, Rutgers, The State University of New Jersey, Piscataway, NJ 08855, USA}

\author{E.~J.~Downie}
\affiliation{Department of Physics, The George Washington University, Washington, DC 20052, USA}

\author{Z.~Duan}
\affiliation{Department of Physics and Astronomy, Rutgers, The State University of New Jersey, Piscataway, NJ 08855, USA}

\author{I.~P.~Fernando}
\altaffiliation{Current: Department of Physics, University of Virginia, Charlottesville, VA 22904, USA}
\affiliation{Department of Physics, Hampton University, Hampton, VA 23668, USA}

\author{A.~Flannery}
\altaffiliation{Current: Department of Physics, Hampton University, Hampton, VA 23668, USA}
\affiliation{Department of Physics and Astronomy, University of South Carolina, Columbia, SC 29208, USA}

\author{D.~Ghosal}
\altaffiliation{Current: Cockcroft Institute, Sci-Tech Daresbury, University of Liverpool, Warrington WA4 4AD, UK}
\affiliation{Department of Physics, University of Basel, 4056 Basel, CH}

\author{A.~Golossanov}
\affiliation{Department of Physics, The George Washington University, Washington, DC 20052, USA}

\author{J.~Guo}
\affiliation{Department of Physics and Astronomy, Rutgers, The State University of New Jersey, Piscataway, NJ 08855, USA}

\author{N.~S.~Ifat}
\affiliation{Department of Physics, Temple University, Philadelphia, PA 19122, USA}

\author{Y.~Ilieva}
\affiliation{Department of Physics and Astronomy, University of South Carolina, Columbia, SC 29208, USA}

\author{M.~Kohl}
\affiliation{Department of Physics, Hampton University, Hampton, VA 23668, USA}

\author{I.~Lavrukhin}
\affiliation{Department of Physics, The George Washington University, Washington, DC 20052, USA}
\affiliation{Randall Laboratory of Physics, University of Michigan, Ann Arbor, MI 48109, USA}

\author{L.~Li}
\altaffiliation{Current: ASML, 80 W Tasman Dr, San Jose, CA 95134, USA}
\affiliation{Department of Physics and Astronomy, University of South Carolina, Columbia, SC 29208, USA}

\author{W.~Lorenzon}
\affiliation{Randall Laboratory of Physics, University of Michigan, Ann Arbor, MI 48109, USA}

\author{P.~Mohanmurthy}	
\altaffiliation{Current: Laboratory for Nuclear Science, Massachusetts Institute of Technology, Cambridge, MA 02139, USA}
\affiliation{Department of Physics, University of Chicago, Chicago, IL 60615, USA}

\author{S.~J.~Nazeer}
\affiliation{Department of Physics, Hampton University, Hampton, VA 23668, USA}

\author{M.~Nicol}
\affiliation{Department of Physics and Astronomy, University of South Carolina, Columbia, SC 29208, USA}

\author{T.~Patel}
\affiliation{Department of Physics, Hampton University, Hampton, VA 23668, USA}

\author{A.~Prosnyakov}
\affiliation{Racah Institute of Physics, Hebrew University of Jerusalem, Jerusalem 91904, IL}

\author{R.~D.~Ransome}
\affiliation{Department of Physics and Astronomy, Rutgers, The State University of New Jersey, Piscataway, NJ 08855, USA}

\author{R.~Ratvasky}
\affiliation{Department of Physics, The George Washington University, Washington, DC 20052, USA}

\author{H.~Reid}
\affiliation{Randall Laboratory of Physics, University of Michigan, Ann Arbor, MI 48109, USA}

\author{P.~E.~Reimer}
\affiliation{Physics Division, Argonne National Laboratory, Lemont, IL 60439, USA}

\author{R.~Richards}
\affiliation{Department of Physics, Hampton University, Hampton, VA 23668, USA}

\author{G.~Ron}
\affiliation{Racah Institute of Physics, Hebrew University of Jerusalem, Jerusalem 91904, IL}

\author{O.~M.~Ruimi}
\affiliation{Racah Institute of Physics, Hebrew University of Jerusalem, Jerusalem 91904, IL}
\affiliation{Helmholtz Institute, Johannes Gutenberg University Mainz, 55128 Mainz, DE}

\author{K.~Salamone}
\affiliation{Center for Frontiers in Nuclear Science, Stony Brook University, Stony Brook, NY 11794, USA}

\author{N.~Sparveris}
\affiliation{Department of Physics, Temple University, Philadelphia, PA 19122, USA}

\author{N.~Wuerfel}
\affiliation{Randall Laboratory of Physics, University of Michigan, Ann Arbor, MI 48109, USA}
\affiliation{Laboratory for Nuclear Science, Massachusetts Institute of Technology, Cambridge, MA 02139, USA}

\author{D.~A.~Yaari}
\affiliation{Racah Institute of Physics, Hebrew University of Jerusalem, Jerusalem 91904, IL}

% \author{W. Lin}
% \email{win.l@rutgers.edu}
%  \affiliation{Department of Physics and Astronomy, Rutgers,
% The State University of New Jersey, Piscataway, NJ, 08855, USA}
% \author{T. Rostomyan}%
%  \email{Tigran.Rostomyan@psi.ch}
% \affiliation{Paul Scherrer Institute, Villigen, CH-5232, Switzerland}